\documentclass[preprint, authoryear, 11pt]{elsarticle}
\usepackage{color}

\usepackage{amsmath}

\usepackage{amssymb}
\usepackage{url}

\biboptions{authoryear}

\usepackage{latexsym,color}

\newcommand{\be}{\begin{equation}}
\newcommand{\ee}{\end{equation}}
\newcommand{\bi}{\begin{itemize}}
\newcommand{\ei}{\end{itemize}}
\newcommand{\ben}{\begin{enumerate}}
\newcommand{\een}{\end{enumerate}}

\newcommand{\bea}{\begin{eqnarray}}
\newcommand{\eea}{\end{eqnarray}}

\newcommand{\commentout}[1]{{}}

\begin{document}
   
\begin{frontmatter}




\title{Human mobility and time spent at destination: impact on spatial epidemic spreading}


\author[epilab,INSERM,UPMC]{Chiara Poletto}
\author[epilab]{Michele Tizzoni}

\author[INSERM,UPMC,isi]{Vittoria Colizza\corref{cor1}\fnref{fn1}}
\ead{vittoria.colizza@inserm.fr}

\address[epilab]{Computational Epidemiology Laboratory, Institute for
  Scientific Interchange (ISI) Foundation, Turin, Italy}
  \address[INSERM]{INSERM, U707, Paris, France}
  \address[UPMC]{UPMC UniversitŽ\'e Paris 06, FacultŽ\'e de MŽ\'edecine Pierre et Marie Curie, UMR S 707, Paris, France}
 \address[isi]{Institute for Scientific Interchange (ISI) Foundation, Turin, Italy}
 
\cortext[cor1]{Corresponding author}


\begin{abstract}

Host mobility plays a fundamental role in the spatial spread of infectious diseases. 
Previous theoretical works based on the integration of network theory into the metapopulation framework have shown that the heterogeneities that characterize real mobility networks favor the propagation of epidemics. 
Nevertheless, the studies conducted so far assumed the mobility process to be either Markovian (in which the memory of the origin of each traveler is lost) or non-Markovian with a fixed traveling time scale (in which individuals travel to a destination and come back at a constant rate). 
Available statistics however show that the time spent by travelers at destination is characterized by wide fluctuations, ranging between a single day up to several months. 
Such varying length of stay crucially affects the chance and duration of mixing events among hosts and may therefore have a strong impact on the spread of an emerging disease. 
Here, we present an analytical and computational study of epidemic processes on a complex subpopulation network where travelers have memory of their origin and spend a heterogeneously distributed time interval at their destination. 
Through analytical calculations and numerical simulations we show that the heterogeneity of the length of stay alters the expression of the threshold between local outbreak and global invasion, and, moreover, it changes the epidemic behavior of the system in case of a global outbreak.
Additionally, our theoretical framework allows us to study the effect of changes in the traveling behavior in response to the infection, by considering a scenario in which sick individuals do not leave their home location.
Finally, we compare the results of our non-Markovian framework with those obtained with a classic Markovian approach and find relevant differences between the two, in the estimate of the epidemic invasion potential, as well as of the timing and the pattern of its spatial spread. 
These results highlight the importance of properly accounting for host trip duration in epidemic models and open the path to the inclusion of such additional layer of complexity to the existing modeling approaches.

\end{abstract}


\end{frontmatter}


\section{Introduction}

The spatial distribution of hosts and their mobility behavior represent two key ingredients in the spatial dissemination of an infectious disease affecting the host population. Modeling approaches that take into account these ingredients can  address crucial epidemiological issues on the expected outcome, such as e.g. the persistence of an infection in the population or the conditions for the invasion of an emerging epidemic~\citep{Riley2007}.

An  ideal theoretical framework to capture the effects of the spatial structure of a population and to explore its epidemiological implications is given by the metapopulation approach~\citep{Hanski1997, Grenfell1997, Tilman1997, Bascompte1998, Hanski2004}. 
This framework has been widely used in population ecology and epidemiology and describes the dynamics of a population in a fragmented environment where a discrete number of localized subpopulations or patches are connected by mobility fluxes. 
In metapopulation epidemic models, individuals belong to well-defined social or geographical units (e.g. households, towns,  or large urban areas) and the coupling among these units is generated by the mobility connections and determines the disease circulation on the spatial system~\citep{Hethcote1978, May1984, Bolker1995, Sattenspiel1995, Keeling2002, Grenfell1997, Ferguson2003}. 

Recently, metapopulation models have been integrated with empirical data on human demography and mobility to create data-driven computational tools for the analysis of large-scale geographic spread of infectious diseases~\citep{Grais2004, Hufnagel2004, Colizza2006a, Cooper2006,Colizza2007a, Epstein2007,Balcan2009b}. 
The use of real data on human mobility has uncovered the important role of the various heterogeneities that characterize human movement patterns on the resulting epidemic. The network structure of human patterns is usually defined as {\em complex}, indicating that it displays a large variability, spanning several orders of magnitude, in the number of connections between locations and in the number of travelers between each origin and destination centre. Such complex features have been found in the worldwide air travel patterns~\citep{Barrat2004, Guimera2005} and in commuting patterns between urban areas~\citep{Chowell2003, Brockmann2006, Gonzalez2008,  Balcan2009b}. 
These fluctuations are particularly relevant when human mobility networks are used as a substrate for metapopulation epidemic models. 
Under very general assumptions, it has been shown that a metapopulation system may be characterized by two epidemic thresholds: a local epidemic threshold, which regulates the spread of the infection within a single subpopulation~\citep{Anderson1992}, and a global epidemic threshold that determines if the epidemic invasion can reach a significant fraction of subpopulations~\citep{Ball1997, Cross2005, Colizza2007b, Colizza2007c,Colizza2008}. The latter distinguishes cases of spatial invasion from  outcomes in which, despite an ongoing outbreak in the seed subpopulation, the epidemic is not able to spread spatially 
because of a small enough mobility rate, which does not ensure the travel of infected individuals to other subpopulations before the end of the local outbreak, or which produces small enough seeding events not enabling the  start of an outbreak in the reached subpopulation due to local extinction events.
Assuming that individuals are homogeneously mixed within each subpopulation, the local threshold depends on the disease parameters only, but the global threshold depends also on the statistical fluctuations of the network connectivity and on the mobility fluxes. 
In case of a broad distribution of connections per subpopulation, while the condition for the occurrence of the local outbreak
remains unchanged, the topological fluctuations lower the threshold condition for the global invasion, thus strongly favoring the spatial spread of the epidemic~\citep{Colizza2007b, Colizza2007c,Colizza2008}. 

Within the metapopulation framework, such results were obtained under the assumption that the mobility of individuals and the concurrent epidemic process can be modeled as particle reaction-diffusion processes~\citep{Colizza2007c}. The first studies also assumed a Markovian dynamics, representing  individuals who are indistinguishable regarding their travel pattern, so that at each time step the same travelling probability applies to all individuals without having memory of their origin~\citep{Colizza2007b, Colizza2007c,Colizza2008}. This assumption has also been adopted in several modeling approaches for the data-driven  large-scale spreading of infectious diseases~\citep{Rvachev1985,Longini1988,Flahault1991,Grais2003,Grais2004,Hufnagel2004,Cooper2006,Epstein2007,Colizza2007a,Colizza2007d,Balcan2009b}, mainly for simplification purposes and absence of exhaustive detailed origin-destination data. Human mobility has however a clear territorial nature characterized, on average, by individuals spending short periods away of their permanent location. In addition, recent results on the analysis of  detailed mobility data at the individual level have  pointed out the high level of predictability and recurrence of individual daily travel patterns~\citep{Wang2009b, Song2010a, Song2010b}. Such recurrent patterns have therefore been included in spatially structured approaches with a specific focus on commuting modes of mobility, that is the type of recurrent daily mobility from the location of residence to the location of work~\citep{Sattenspiel1995,Danon2009, Keeling2010, Balcan2011, Balcan2012, Belik2011}. This corresponds to recording the subpopulation of residence for each individual, assuming that they spend a fixed time at destination, before returning to their residence at each timestep. 

\begin{figure*}[th!]
\begin{center}
\includegraphics[width=14cm]{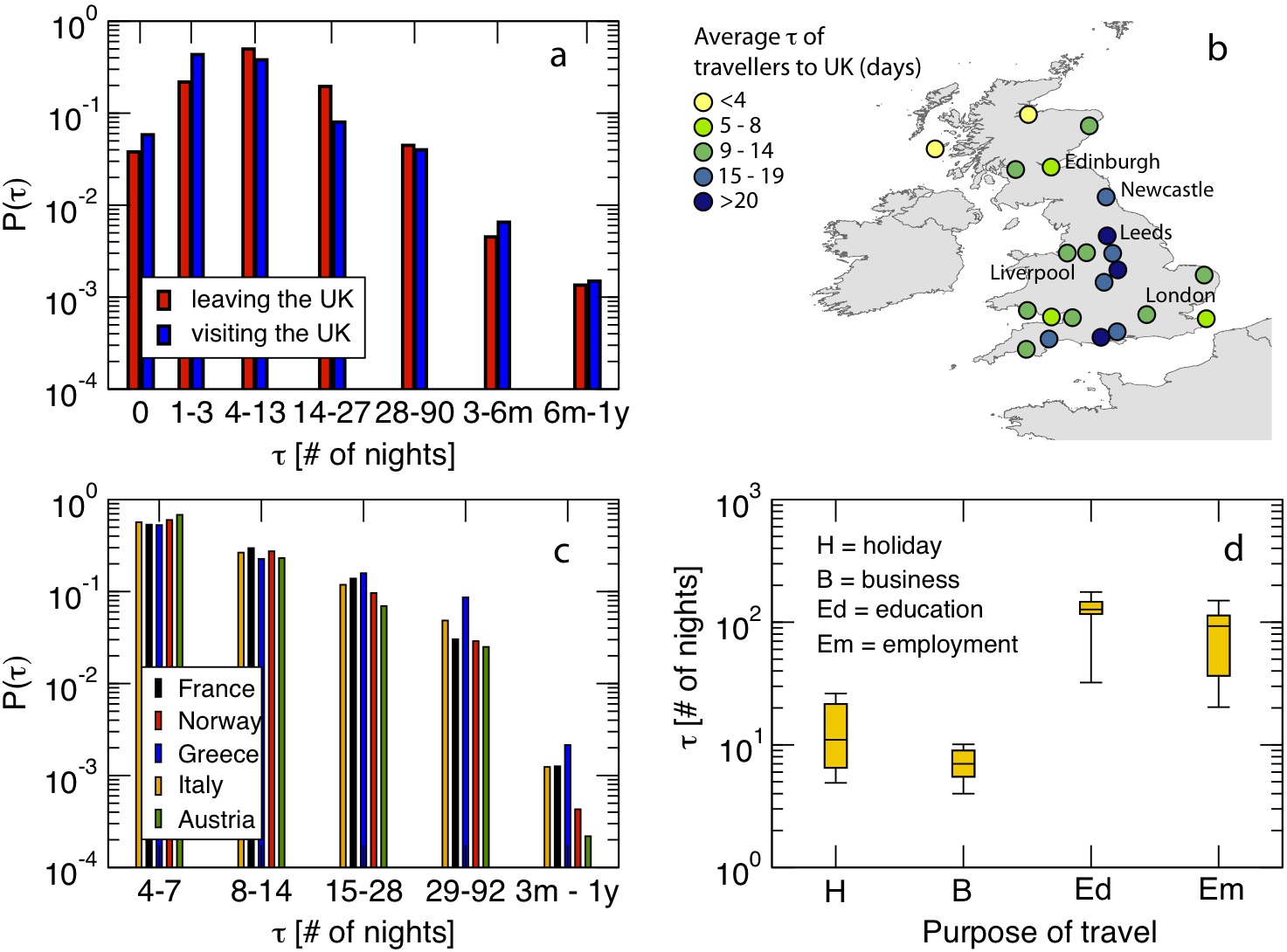} 
\caption{Empirical data on the length of stay of travelers. $(a)$ The number of nights spent by foreign travelers visiting the UK (in blue) and by British traveling abroad (in red) spans several orders of magnitude, from 1 night to several months, considering all travel purposes and all countries of origin and destination (source: UK Travel Trends 2009). $(b)$ Geographic variations in the average length of stay of foreign travellers in different cities of the UK (source: UK Travel Trends 2009). $(c)$ The number of nights spent by tourists traveling to European countries on holiday is broadly distributed. Here we show data for 5 selected countries of destination, considering only trips of 4 nights or more (source: Eurostat). $(d)$ Box plot of the length of stay, measured as number of nights, of travelers who visited Australia in 2008 according to the purpose of their visit: holiday, business, educational and employment. The statistics is obtained over a sample of 19 countries of origin (source: Tourism Research Australia).\label{fig:data}
}
\end{center}
\end{figure*}

Additional modes of mobility, besides the commuting behavior, are characterized by variable lengths of stay at destination, mainly dependent on the purpose of the trip but also on its logistical details and the accessibility of the destinations. Data collected from the office of statistics of several countries worldwide indicate that the time spent by travelers at their destination is broadly distributed and ranges from a single day to months. 
In Fig.~\ref{fig:data} we present some examples that support such empirical evidence for a number of countries. The number of nights spent by foreign travelers in the UK and other European countries span several orders of magnitude, ranging from less than a week to 6 months or more, considering all countries of origins and all travel purposes.
Importantly, the average time spent by tourists in different British cities shows strong geographical variations: some locations are characterized by a short length of stay (less than 4 days on average) while in other cities travelers spend on average more than 20 days.
The length of stay also depends strongly on the purpose of the trip. As shown by Australian data, business trips tend to be shorter, usually lasting less than 15 days, while educational and employment trips are usually very long and can easily exceed the 3 months duration, on average. Holiday trips show large fluctuations in their duration and can last from few days up to 50 days, on average.
Furthermore, similar large fluctuations in the duration of visits were also observed on different time and spatial scales. For instance, the time spent by residents in different locations within a US city during their daily routine varies from few minutes to several hours~\citep{Eubank2004, Eubank2006}. Similarly, the analysis of a large dataset of mobile phone records revealed that the time spent by mobile phones users at a given location identified by a phone tower cell, $\Delta t$, follows a power law distribution $P(\Delta t) \sim |\Delta t|^{-(1+\beta)}$ with $0 < \beta \leq 1$~\citep{Gonzalez2008, Song2010b}.

Prompted by these empirical findings, we study spreading processes on a metapopulation system where individuals have memory of their subpopulation of origin and spend at destination an average length of stay $\tau$ that is broadly distributed.  This corresponds to focus on an additional level of heterogeneity that is observed in human mobility and that is associated to the timescales of travel movements.

In order to integrate the large fluctuations of $\tau$ observed in reality into our modeling framework, we assume that the length of stay of travelers depends on their destination only and it varies from place to place, as a power-law function of the connectivity of a subpopulation, similar to the degree dependence observed in other metapopulation variables (such as e.g. the population size, the total flux of travelers, and others)~\citep{Poletto2012}. 
Following~\citep{Poletto2012}, we show analytically that, similar to what is observed with a Markovian dynamics, the system is characterized by a global epidemic threshold which defines a transition between a regime where only few subpopulations are affected by the spreading process and a regime where the infectious agent spreads globally on the network. 
Such a threshold, whose expression can be computed analytically, depends both on the network topology and on the considered distribution of the length of stay. In this paper we fully explore the dependence of the invasion condition on the transportation structure, the travel volumes of passengers and their spatial distribution, and the parameter describing the expression of the time interval spent at destination.
A longer time spent by travelers in peripheral locations, with respect to large hubs, can lead to a substantial suppression of the disease transmission at the global level, and, conversely, the spreading process may be accelerated if a longer length of stay characterizes the busiest locations.
We confirm our analytical results with numerical Monte Carlo simulations on synthetic metapopulation networks, where single individuals are tracked in time and all the modeled processes are fully stochastic.  

In addition to the study of the invasion dynamics and its dependence on the features of the metapopulation system,
we test here the effects of changes in the travel behavior of individuals induced by the illness, exploiting the discrete nature of the model. In particular, we study how the expression of the global threshold changes when ill individuals do not leave home or when individuals who got infected during their journey return to their place of residence and do not travel until they recover. Moreover, we study the behavior of the system above the epidemic threshold, characterizing its dynamics and invasion pattern, and compare the results obtained within the non-Markovian framework with those obtained using a classic Markovian approach. 
Our results provide new insights on the impact of heterogeneous mobility timescales on the spatial spread of infectious diseases and open the path to a more realistic description of mobility processes in epidemic modeling.

The paper is organized as follows: Section~\ref{sec:metapop} introduces the formalism for metapopulation epidemic models. In particular, we consider the case of a non-Markovian traveling dynamics with a substrate mobility network characterized by a heterogeneous distribution of links per subpopulation and by heterogeneous traveling fluxes, as observed in reality. We also consider the large fluctuations associated to the length of stay of travelers at destination in such movement dynamics. 
Section~\ref{sec:global_th} integrates the modeling of the disease spreading on the metapopulation system. Assuming that the invasion dynamics at the level of subpopulation can be described as a branching process, we show the existence of an invasion threshold for the metapopulation system and compute an explicit expression of it, discussing its dependence on the network structure and the features of the length of stay~\citep{Poletto2012}. Analytical results are validated through a  numerical analysis confirming the dependence of the invasion condition on the various aspects characterizing the system found in the analytical results. We further extend the framework first introduced in~\citep{Poletto2012} and study how changes in the travel behavior following illness affect the global threshold of the system (Section~\ref{sec:oneseed}). We then analyze the system behavior above the threshold, showing how different distributions of the length of stay result in different paths of infection (Section~\ref{sec:above_th}), and conclude with a comparison to the Markovian dynamics (Section~\ref{sec:mark_nomark}).

\section{Metapopulation model with heterogeneous length of stay\label{sec:metapop}}

\begin{figure*}[t]
\begin{center}
\includegraphics[width=9cm]{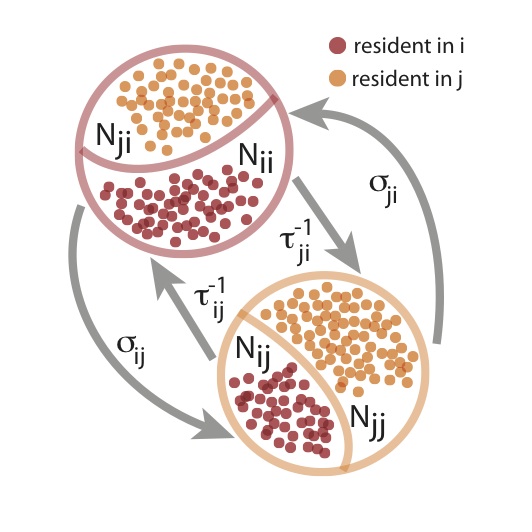} 
\caption{At any time any subpopulation $i$ is occupied by a fraction of its own population $N_{ii}$ and a fraction of individuals $N_{ji}$ resident in the neighboring subpopulation $j$ and who are currently visiting the $i$ subpopulation. Travelling individuals from $i$ leave their home subpopulation to the subpopulation $j$ with rate $\sigma_{ij}$ and return back with rate $\tau^{-1}_{ij}$, where $\tau_{ij}$	is the average time spent at destination. Here we assume that the length of stay is function of the destination only, namely  $\tau_{ij} \equiv  \tau_{j}$.\label{fig:non-markov-scheme}}
\end{center}
\end{figure*}

In order to study the effects of a non-Markovian dynamics with a heterogeneously distributed length of stay on the epidemic spread, we consider a metapopulation system with $V$ subpopulations connected by edges, representing travel connections along which individuals can migrate. Each subpopulation $i$ is characterized by $N_i$ residents ($i=1 \dots V$) and it is connected to a set of $v(i)$ other subpopulations. Since we want to keep track of the origin and destination of travelers during the mobility process, we define the class $N_{ij}$ of individuals resident in $i$ and present in $j$, along with the class $N_{ii}$ of those who are resident in $i$ and located in $i$ -- see the model scheme in Fig.~\ref{fig:non-markov-scheme}. We assume that individuals leave their origin subpopulation $i$ to visit subpopulation $j$ with a per capita diffusion rate $\sigma_{ij}$ and spend at their destination a time interval $\tau_{ij}$, the so called length of stay, before returning home. The length of stay accounts for the timescale characteristic of the trip. In principle this timescale may depend on both origin and destination, but in the following we assume that it depends only on the trip destination, therefore $\tau_{ij} \equiv \tau_j$.

The mobility dynamics is fully described by the following set of equations:
\begin{eqnarray}
\partial_t N_{ii}(t) &=& - \sum_{l \in v(i)} \sigma_{il} N_{ii}(t) + \sum_{l \in v(i)} N_{il}(t)/\tau_{l} \label{eq:no-mark1}\\
\partial_t N_{ij}(t) &=& \sigma_{ij} N_{ii}(t) - N_{ij}(t)/\tau_{j}\,\label{eq:no-mark2},
\end{eqnarray}
where the time evolution of the class $N_{ii}$ is determined by the net balance between the flux of people leaving the subpopulation $i$ with total rate $\sigma_i = \sum_{l \in v(i)} \sigma_{il}$ and the flux of travelers coming back from all their destinations $l$. The second equation describes the time evolution of $N_{ij}$ as the difference between  incoming and returning travelers to subpopulation $i$, with rates $\sigma_{i}$ and $\tau^{-1}_{j}$, respectively.
The above equations can be solved, using the following relation for the total population $N_i$, which is constant in time:
\begin{equation}
N_i = N_{ii}(t) + \sum_j N_{ij}(t) \,.
\end{equation}

It is possible to show that the solutions for $N_{ii}(t)$ and $N_{ij}(t)$ are characterized by the relaxation times $(\tau^{-1}_i + \sigma_i)$ and $\tau^{-1}_i$, respectively~\citep{Balcan2012}. 
After an initial transient, the classes $N_{ii}$ and $N_{ij}$ for each subopopulation $i$ and $j$ therefore reach the stationary values:
\begin{equation}
\overline{N}_{ii} = \frac{N_i}{1+ \sigma_i \tau_i} ,\;\; \overline{N}_{ij} = \frac{\sigma_{ij} \tau_i N_i}{1+ \sigma_i \tau_i}.\label{eq:eq_pop}
\end{equation}
Under the assumption $\sigma_i \ll \tau^{-1}_j$ for all $i$ and $j$,  the timescale of relaxation to equilibrium is given by $\tau_{max} = \max_j \tau_j$~\citep{Keeling2002}. 
This observation is of crucial importance: when the disease dynamics occurs over timescales that are much longer than $\tau_{max}$, then the population dynamics can be considered at equilibrium with respect to the infection dynamics and it is reasonable to assume all population variables to acquire their stationary values.

When the system is composed by a large number of subpopulations with highly heterogeneous connectivity patterns, as observed in real networked systems, the analytical treatment of Eqs.~(\ref{eq:no-mark1}) and (\ref{eq:no-mark2}) becomes unfeasible. 
However, it is possible to adopt a mean-field approach to provide a coarse grained description of the mobility dynamics, which takes into account the system's connectivity pattern only and disregards all other details. Such mean-field approach is called degree-bock approximation and it assumes that nodes with the same degree are statistically equivalent. This approach has been successfully applied to model several dynamical processes on complex networks~\citep{Pastorsatorras2001a,Pastorsatorras2001b,Colizza2007b,Colizza2007c,Colizza2008, Balcan2011,Balcan2012, Meloni2011}, and it is supported by the empirical evidence reporting on the degree dependence of numerous system's variables, like e.g. the population size of each subpopulation, the traffic amount at each node, and others~\citep{Barrat2004,Colizza2006a,Colizza2008}.

We consider a random metapopulation system with a given degree distribution $P(k)$. The subpopulations are divided into classes according to their degree and their properties are described by average quantities through the degree-block approximation: 
\begin{equation}
X_k= \frac{1} {V_k} \sum_{i|k_i= k} X_i ,
\end{equation}
where $V_k$ is the number of subpopulations with degree $k$, $X_i$ the quantity of interest for the subpopulation $i$, and the sum runs over all subpopulations having degree $k_i$ equal to $k$.
Following the empirical evidence, we assume that the population of each node and the number of travelers between two subpopulations are defined by scaling relations with the degree of the nodes. 
In particular, we assume that the population of a node of degree $k$, $N_k$, is given by:
\begin{equation}
N_k = \overline{N} \frac{k^\phi}{\langle k^\phi \rangle},
\end{equation}
where $\overline{N}=\sum_k N_k P(k)$ is the average number of residents in each subpopulation of the system.
Additionally, we assume that the number of individuals migrating between a subpopulation of degree $k$ and a subpopulation of degree $k'$ is defined by:
\begin{equation}
w_{kk'} = w_0 (kk')^\theta \,.
\end{equation}
The exponents $\theta$ and $\phi$, and the scaling factor $w_0$, vary according to the  real system under study. The mobility of individuals along the connections of the network is modelled with the per capita diffusion rate $\sigma_{kk'}=w_{kk'}/N_k$ and the overall leaving rate out of the node with degree $k$ is given by $\sigma_k=k\sum_{k'}\sigma_{kk'}P(k'|k)$, where $P(k'|k)$ is the conditional probability that a node with degree $k$ is linked to a node with degree $k'$.
We define the leaving rate rescaling as $\sigma=w_0 \langle k^\phi \rangle /\overline{N}$, so that the diffusion rate equation reads
\begin{equation}
\sigma_{kk'}=\sigma k^{\theta-\phi} k'^\theta .
\label{eq:diffusion_rate}
\end{equation}
Finally, travelers spend at their destination characterized by a degree $k'$ an average time $\tau_{k'}$. This means that in our model the length of stay is a specific characteristic of a given location of destination, and it is fully determined by its degree.

Within this framework, we want to explore how different levels of heterogeneity in the distribution of the length of stay affect the epidemic process, modifying the underlying traveling dynamics.
In the following, we assume that the length of stay characterizing a subpopulation with degree $k$ is described by the functional form:
\begin{equation}
\tau_k = \frac{\overline{\tau}}{\langle k^\chi \rangle}k^\chi\,,
\label{eq:tau}
\end{equation}
where $\overline{\tau} = \sum_k \tau_k P(k)$ is the average length of stay over the whole metapopulation network~\citep{Poletto2012}. Different values of the power-law exponent $\chi$ define different dynamical regimes of the system. For $\chi>0$ the length of stay is positively correlated with the degree of the subpopulation of destination, which implies that individuals traveling to a well connected location will spend a longer time at destination than individuals traveling to remote locations. 
In this regime, locations that are important for their socio-economic, touristic or geographic characteristics, thus  being easily accessible through  a large number of connections, are considered also attractive in terms of the amount of time spent by each visitor at these destinations.  An example on a large geographical scale is represented by large urban areas served by airport hubs, where an efficient connectivity in terms of transport also corresponds to large attractiveness of the location for tourism or seasonal/temporary job opportunities~\citep{Lohmann2009}.
The opposite regime is defined by $\chi<0$, where the length of stay increases as the connectivity of the subpopulation of destination decreases. In this case, since small degree locations are usually peripheral in the transportation system, a longer length of stay may be explained by an optimization choice made by the traveler between the time spent at destination and the time spent to reach the destination~\citep{McKercher2003}. 
The value $\chi=0$ corresponds to the case of homogeneous length of stay, as it is generally assumed for  commuting-like process, where the length of stay $\tau$ represents the duration of an average working day~\citep{Belik2011, Balcan2011, Balcan2012}. 
On the other hand, the limit $\chi\rightarrow \infty$ corresponds to the case of permanent migration, which is commonly used 
to model mobility processes with no return to the origin (such as the case of livestock displacements in
trade flows~\citep{Bajardi2011b,Bajardi2012}) or to approximate origin-destination mobility by
simplifying the modelling approach and assuming a Markovian process\citep{Rvachev1985, Colizza2007c, Balcan2009a}.

Here we want to focus on a range of values for the parameter $\chi$, that can well capture the empirically observed heterogeneity of the length of stay. 
Our assumption that the length of stay depends on the geographic location of a node, and is fully encoded by its degree $k$, is also based on the empirical evidence reported in Fig.~\ref{fig:data}. Obviously, this is a simple assumption that we consider since it allows us to analytically solve the epidemic metapopulation model. The expression of $\tau$ may be modified to include additional factors that may affect the length of stay, such as the distance between locations or other socio-economic indicators. Though representing  a fundamental ingredient for the modeling of a visitor's decision-making process,  
there is no clear consensus in the economic literature on the determinants of the length of stay and the matter is still largely debated~\citep{Gokovali2007}. Therefore, alternative choices to our assumption cannot be currently selected solely based on the available knowledge of the problem. Our perspective is to show that, even under simple assumptions, the heterogeneity of the mobility timescales modifies the system's behavior and introduces a number of new and interesting dynamical properties that impact the epidemic spread.

By considering the assumptions illustrated above, we reformulate the non-Markovian mobility dynamics within the degree-block description. The mobility process described by Eq.~(\ref{eq:no-mark1}) and (\ref{eq:no-mark2}) can be translated into the degree block notation by defining all the population variables in terms of the subpopulation's degree.  Each subpopulation of degree $k$ has $N_k$ inhabitants. They are further divided into two classes: those who are from $k$ and are located in $k$ at time $t$, $N_{kk}(t)$, and those who are from $k$ and are located in a neighboring subpopulation of degree $k'$ at time $t$, $N_{kk'}(t)$. 
The resulting equations that describe the non-Markovian travelling dynamics are:
\begin{eqnarray}
\partial_t N_{kk}(t) &=& - \sigma_{k} N_{kk}(t) + k \sum_{k'} N_{kk'}(t) P(k'|k) \tau^{-1}_{k'} \label{eq:no-mark1-deg}\\
\partial_t N_{kk'}(t) &=& \sigma_{kk'} N_{kk}(t) - N_{kk'}(t) \tau^{-1}_{k'}\,\label{eq:no-mark2-deg}.
\end{eqnarray}
The condition $\partial_t N_{kk}(t)=\partial_t N_{kk'}(t)=0$ yields the equilibrium solutions:
\begin{equation}
\overline{N}_{kk} = \frac{\overline{N}}{\langle k^\phi \rangle} \nu_k k^\phi ,
\label{eq:pop0-deg}
\end{equation}
\begin{equation}
\overline{N}_{kk'} = \frac{\sigma \overline{N} \overline{\tau}}{\langle k^\phi \rangle \langle k^\chi \rangle} \nu_k k^\theta k'^{\theta + \chi} \label{eq:pop1-deg}.
\end{equation}
as detailed in \ref{app:populations}. In the above expressions, the factor $\nu_k$ is given by:
\begin{equation}
\nu_{k} = \left( 1 + \sigma \overline{\tau} \frac{\langle k^{\theta + \chi +1} \rangle}{\langle k \rangle \langle k^\chi \rangle} k^{\theta - \phi +1} \right)^{-1} \label{eq:pop2-deg}.
\end{equation}
Adopting the timescale separation approximation originally proposed in~\citep{Keeling2000, Keeling2002}, we can consider the equations (\ref{eq:pop1-deg}) and (\ref{eq:pop2-deg}) to be a good approximation to the system behavior when $\sigma_k \ll \tau^{-1}_k$.

\section{Global Invasion Threshold \label{sec:global_th}}
We model the epidemic process within each subpopulation by considering a standard susceptible-infectious-recovered (SIR) model with transmission rate $\beta$ and recovery rate $\mu$~\citep{Anderson1992}. Recovered individuals are considered permanently immune to the disease. 
The occupation number of each compartment in the subpopulation $i$, $X_i$ (where $X  = S, I$ or $R$), is divided into the subclasses $X_{ij}$ and $X_{ii}$ corresponding to the individuals in the disease state $X$ who are resident in $i$
and are located in subpopulation $j$ or $i$, respectively~\citep{Keeling2002,Sattenspiel1995}. 
We assume homogeneous mixing within each subpopulation, therefore all susceptible individuals who are present in $i$ at a given time step are subject to the same force of infection $\mathcal F_i(t)= \beta I_i^*(t)/N_i^*(t)$ defined by the mass-action principle. $\mathcal F_i(t)$ depends on the number of infectious individuals simultaneously present in $i$, $I_i^*(t)$,  and the temporary population formed by resident and non-resident individuals, $N_i^*(t)$.

In a fully susceptible population, the ratio $R_0 = \beta/\mu$ defines the basic reproduction number, that is the average number of secondary cases generated by an infectious individual during his infectious period~\citep{Anderson1992}. When $R_0>1$ the epidemic will affect a non-negligible fraction of the population of the seeded subpopulation, thus representing a threshold for a local epidemic.
In a spatially extended system, however, the condition on $R_0$ is not sufficient to guarantee the propagation of the epidemic out of the initial seed to reach a finite fraction of the whole system. This may happen because of small enough flows of travelers that are not able to allow the diffusion of infectious individuals before the outbreak dies out in the seed, or because of local extinction events in newly infected subpopulation due to rare or limited seeding events. Next to the local threshold condition, an additional predictor is therefore needed to define the global invasion threshold in a metapopulation system, $R_*$, that governs the disease transmission between subpopulations and depends both on the epidemic features and on the mobility parameters describing individuals' movements~\citep{Ball1997,Cross2005,Colizza2007b,Colizza2007c,Colizza2008}. The time length spent at destination by each individual is clearly an important factor that may affect the conditions for the global threshold, in that it represents the time during which travelers may be exposed to an outbreak in a trip to an affected area or during which the passengers themselves, carrying the infection, may transmit the disease to the population at destination~\citep{Poletto2012}. In the following subsection we will explore in detail the role of the length of stay at destination and of its fluctuations on the global invasion threshold and on the epidemic dynamics.

\subsection{Analytical treatment}
Following the approach of~\citep{Colizza2007c,Colizza2008,Balcan2012}, we describe the disease invasion at the subpopulation level using a branching tree approximation~\citep{Harris1989,Ball1997,Vazquez2006}, where each subpopulation is treated as the basic elements of the process. 
We assume that a local outbreak is taking place in a given subpopulation of degree $k$ and then we follow the spread of the infection from that subpopulation to the others by means of the underlying mobility network. 
Each subpopulation is defined as not infected if no outbreak is taking place in it, or \emph{diseased} otherwise.
Then, the invasion process starts from an initial set of diseased subpopulations of degree $k$, $D^0_k$ and each of them infects some of its neighbors, leading to a second generation of diseased subpopulations, $D^1_k$.
We can generalize the notation indicating with $D^n_k$ the number of diseased subpopulations of degree $k$ at generation $n$ and derive the relation between subsequent generations of diseased subpopulations, $D^{n}_k$ and $D^{n-1}_k$.
Assuming that the mobility network is uncorrelated and that the value of $R_0$ is slightly exceeding the epidemic threshold, $R_{0}-1 \ll 1$, it is possible to show that:
\begin{equation}
D^n_k=(R_0-1) \frac{kP(k)}{\langle k \rangle} \sum_{k'}D^{n-1}_{k'} (k'-1) \lambda_{k'k} \,,
\label{eq:branching}
\end{equation}
where $\lambda_{k'k}$ represents the number of infectious individuals that can travel from a diseased subpopulation of degree $k'$ to a non-diseased subpopulation of degree $k$ (see details in \ref{app:threshold}). 

The latter term is the one that relates the microscopic dynamics of the local infection taking place within a subpopulation to the coarse-grained view that describes the disease invasion at the metapopulation level. 
Given a diseased subpopulation of degree $k$ connected to a disease-free subpopulation of degree $k'$, the number of possible seeders is the sum of infectious individuals resident in $k$ and traveling to $k'$ and susceptible individuals of $k'$ traveling to $k$, catching the disease and coming back infected. 
Indicating with $\alpha$ the proportion of individuals that will experience the disease by the end of the epidemic in a given population, and that can be approximated by the standard SIR attack rate equation $\alpha \simeq 2(R_0-1)/R_0$ for $R_0 \simeq 1$, the quasi-equilibrium approximation implies that the traveling individuals $N_{kk'}$ will be infected in the same proportion. 
This holds also for the individuals of $k'$ visiting $k$, since they are subject to the same force of infection of the residents in $k$. 
Therefore, the quantity $\lambda_{k'k}$ can be expressed as:
\begin{equation}
\lambda_{k'k} = (\overline{N}_{kk'} + \overline{N}_{k'k})\alpha \,,
\label{eq:2seeds}
\end{equation}
where we used the stationary values of Eq.~(\ref{eq:pop1-deg}), adopting a timescale separation approximation~\citep{Keeling2002,Balcan2009b,Balcan2011}. 
This approach is based on the assumption that the timescale associated to the disease is much larger than the timescale associated to the mobility process, $\mu^{-1} \gg \tau_k$, for any degree $k$.
It is worth to notice that Eq.~(\ref{eq:2seeds}) holds under the assumption that all individuals can travel, regardless of their disease state. It is however important to consider that in reality infectious symptomatic individuals will generally have a reduced probability of traveling, depending on their symptoms and health conditions. For this reason we explore in Section~\ref{sec:oneseed} the effects of changes in the travel behavior due to illness, modifying the expression of Eq.~(\ref{eq:2seeds}).

Plugging the explicit expressions of  Eq.~(\ref{eq:pop1-deg}) into Eq.~(\ref{eq:branching}) and solving the iterative equation as detailed in \ref{app:threshold}, we find that the epidemic dynamic at the global level is ruled by the predictor~\citep{Poletto2012}: 
\begin{equation}
R_* = \frac{2(R_0-1)^2}{R^2_0} \sigma \overline{N} \overline{\tau} \frac{\Lambda({P(k)}, \sigma, \overline{\tau}, \overline{N})}{\langle k \rangle \langle k^\phi \rangle \langle k^\chi \rangle}\,,
\label{eq:threshold}
\end{equation}
where the quantity
\begin{multline}
\Lambda({P(k)}, \sigma, \overline{\tau}, \overline{N}) = \langle (k-1 )k^{2\theta+\chi+1} \nu_k \rangle + \\ \sqrt{\langle (k-1) k^{2(\theta+\chi)+1} \rangle \langle(k-1)k^{2\theta+1} \nu^2_k\rangle}\,,
\label{eq:lambda}
\end{multline}
is a function of the moments of the degree distribution, of the average length of stay $\overline{\tau}$, of the leaving rate rescaling $\sigma$ and of the average subpopulation size $\overline{N}$.

\begin{figure*}[tp]
\begin{center}
\includegraphics[width=9cm]{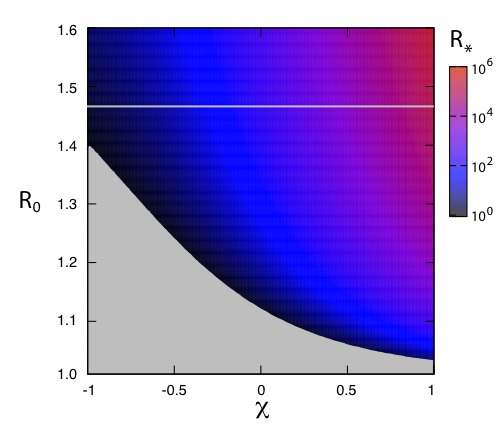} 
\caption{Global reproductive number as a function of $R_0$ and $\chi$ in heterogeneous and homogeneous substrate networks. The Fig.~codes with color the value of the function $R_*(R_0, \chi)$ in the heterogeneous network. The grey area corresponds to the containment region. The solid grey horizontal line indicates the threshold that separates the containment region (below) from the invasion region (above) in the homogeneous network. Both networks are characterized by $V=10^4$ nodes, average degree $\langle k \rangle = 3$ and average length of stay $\overline{\tau}= 37$. The heterogeneous network has a power law degree distribution, $P(k) \propto k^{\gamma}$ with $\gamma= 3$. In the homogeneous network all nodes have the same degree $\overline{k}=3$. The mobility fluxes are characterized by the following parameters: $\sigma= 10^{-5}$, $\phi= \frac 3 4$ and $\theta= \frac 1 2$.\label{fig:th_th}
}
\end{center}
\end{figure*}

The condition $R_*>1$ assures that an infection seeded in a single subpopulation will spread globally and reach a finite fraction of the subpopulations in the metapopulation system. 
Therefore, by solving the equation $R_*=1$ we can find the threshold values for the model's parameters related to the mobility, the demography and the underlying network topology.
Since our main focus is on the effects of the heterogeneous mobility timescales, we look at the dependence of $R_*$ on the parameters $R_0$,  regulating the local epidemic threshold, and $\chi$,  tuning the heterogeneity of the length of stay (Fig.~\ref{fig:th_th} for the case of a heterogeneous substrate network). The latter parameter and its fluctuations strongly affect the behavior of the global threshold. In the case $\chi>0$, the length of stay is positively correlated with the degree of a subpopulation, therefore travelers spend a longer time visiting large hubs, thus enhancing the spreading potential of these locations. The corresponding values of $R_*$ are very large, even in the case of a mild disease. Only for $R_0 \approx 1$ the disease spread can be contained, as indicated by the grey region of the heatmap.  When $\chi<0$, the effect of a heterogeneous topology~--~generally favoring the spread of the disease~\citep{Colizza2007c,Colizza2008}~--~is counterbalanced by a length of stay negatively correlated with the degree of the subpopulation. As visitors tend to spend longer times in peripheral locations, the disease propagation at the global level can be sustained only by larger values of $R_0$ and the containment region becomes wider as $\chi$ decreases.

In order to highlight the role of topological fluctuations, in Fig.~\ref{fig:th_th} we compare the global threshold behavior for a heterogeneous topology with the one obtained for a topologically homogeneous network, where all nodes have the same degree $\overline{k} = 3$. 
In the latter case, the expression of Eq.~(\ref{eq:threshold}) is greatly simplified and it reduces to:
\begin{equation}
R_* =  \frac{4 (R_0-1)^2}{R_0^2} \sigma \overline{N} \overline{\tau} ( \overline{k}-1) \overline{k}^{2 \theta}  \nu_{\overline{k}} \,,
\label{eq:thresh_hom}
\end{equation}
where $\nu_{\overline{k}}  = (1+\sigma \overline{\tau} \overline{k}^{2\theta-\phi+1})^{-1}$, as detailed in \ref{app:homo_threshold}.
It is worth to notice that in the homogeneous network  the length of stay  is constant for all the subpopulations, due to the absence of topological fluctuations, and the exponent $\chi$ disappears from the threshold equation. Therefore, in the $(R_0,\, \chi)$ plane of Fig \ref{fig:th_th} the invasion region is delimited by a constant value of $R_0$, for every $\chi$, as shown by the grey line in the plot.
As it has already been observed for commuting processes or Markovian mobility processes, the topological heterogeneity of the network considerably favors the global disease spread by lowering the threshold value~\citep{Colizza2008, Balcan2012}, while in the case of a homogeneous substrate network the containment region spans a wider range of $R_0$ values.

\begin{figure*}[tp]
\begin{center}
\includegraphics[width=\textwidth]{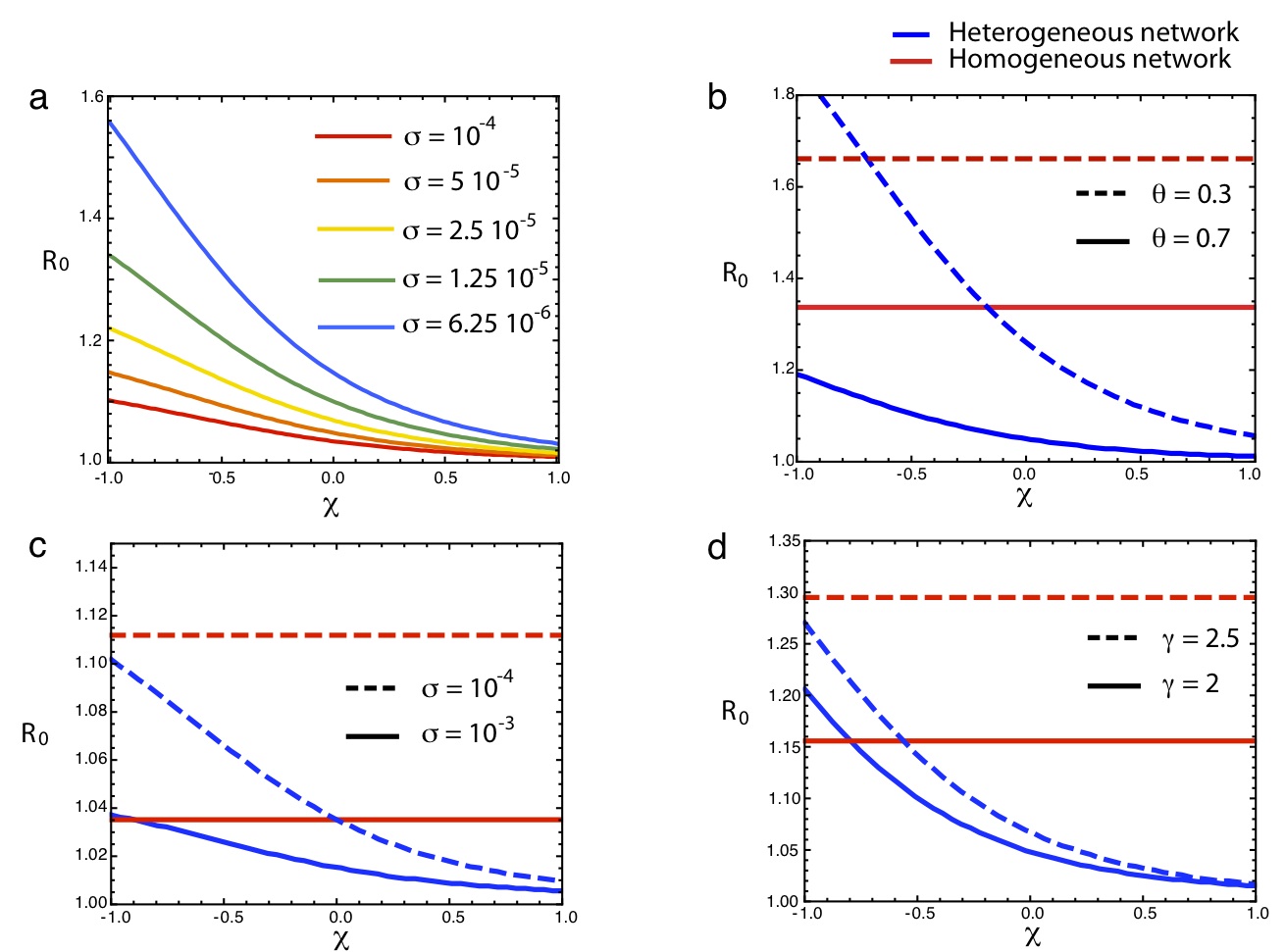} 
\caption{\label{fig:th_th_het-hom_par} Phase diagram defined by the threshold condition $R_*(R_0, \chi)=1$, comparing two network topologies and different values in the model's parameters. Each panel displays the effects of  variations in the parameter indicated in the legend keeping the others the same as in Fig.~\ref{fig:th_th}. In panels b, c and d two different values of the parameter are compared (continuous and dashed lines) for a heterogeneous and a homogeneous network (in blue and red respectively).}
\end{center}
\end{figure*}

The global epidemic threshold is also determined by other ingredients of the metapopulation system; in particular by those related to the architecture of the substrate network, such as the level of heterogeneity of the degree distribution tuned by the exponent  $\gamma$, and those related to the individual mobility patterns, such as the leaving rate rescaling $\sigma$ and the scaling exponent $\theta$. 
In Fig.~\ref{fig:th_th_het-hom_par}, we explore the behavior of the phase diagram of the system under changes in the above mentioned parameters and compare the results obtained with  a heterogeneous network topology to those obtained with a homogeneous one. In all panels, the invasion region $R_*(R_0, \chi)>1$, not colored for the sake of visualization, is always located above the curve, being $R_*$  a monotonous increasing function of $R_0$.
As expected, larger mobility flows between connected subpopulations of a given size, corresponding to larger values of $\sigma$, lower the threshold value.
Indeed, increasing the mobility rate of individuals favors the epidemic spreading and this effect is more relevant when the length of stay is negatively correlated with the degree of a node ($\chi<0$).  In the case $\chi>0$, the spreading potential of the hubs suppresses the effects related to changes in the mobility of individuals leading to a very low threshold for any value of $\sigma$.
These results indicate that, if large hubs are characterized by a shorter length of stay with respect to poorly connected locations, then reducing the mobility of individuals can lead to the containment of an emerging disease. On the other hand, the longer the time spent by travelers in highly connected locations the less effective is any travel restriction measure aimed at containing the global disease spread. A conclusion, although based on a different framework, that is similar to the results on the effectiveness of travel restrictions obtained using a Markovian approach~\citep{Colizza2007b,Bajardi2011a}.
A similar effect is observed by varying the scaling exponent $\theta$: larger values of $\theta$, corresponding to stronger fluctuations in the traffic distribution, enhance the invasion potential of the disease yielding smaller values of the critical basic reproductive number $R_0$ above which a spatial propagation is predicted. On the other hand, a more uniform traffic distribution obtained for $\theta \to 0$ (e.g. $\theta= 0.3$ in the Figure) reduces the spreading potential especially in the case of $\chi<0$. 
For the exponent $\gamma$ that characterizes the degree distribution ($P(k)\sim k^{-\gamma}$) we consider the values 2 and 2.5 corresponding to an increasing degree of topological heterogeneity, respectively. Consistently with previous results, as the topological fluctuations become more relevant the global epidemic threshold becomes smaller.

When comparing two network topologies, we see that, for all the considered changes in the model's parameters, the global threshold is generally lower in the heterogeneous network. 
However, it is worth to notice that in some cases  there is a region in the $(R_0,\, \chi)$ parameter space where the global invasion phase is predicted to occur on a homogeneous topology but not on a heterogeneous one (Fig.~\ref{fig:th_th_het-hom_par}).  
This happens for large and negative values of $\chi$ and indicates that as travelers spend more time in poorly connected locations and less time in large hubs, the topological fluctuations of the network become less relevant and the disease spread may be contained even on a system characterized by a heterogeneous topology.

\subsection{Numerical validation and assessment of analytical assumptions}

\begin{figure*}[tp]
\begin{center}
\includegraphics[width=9cm]{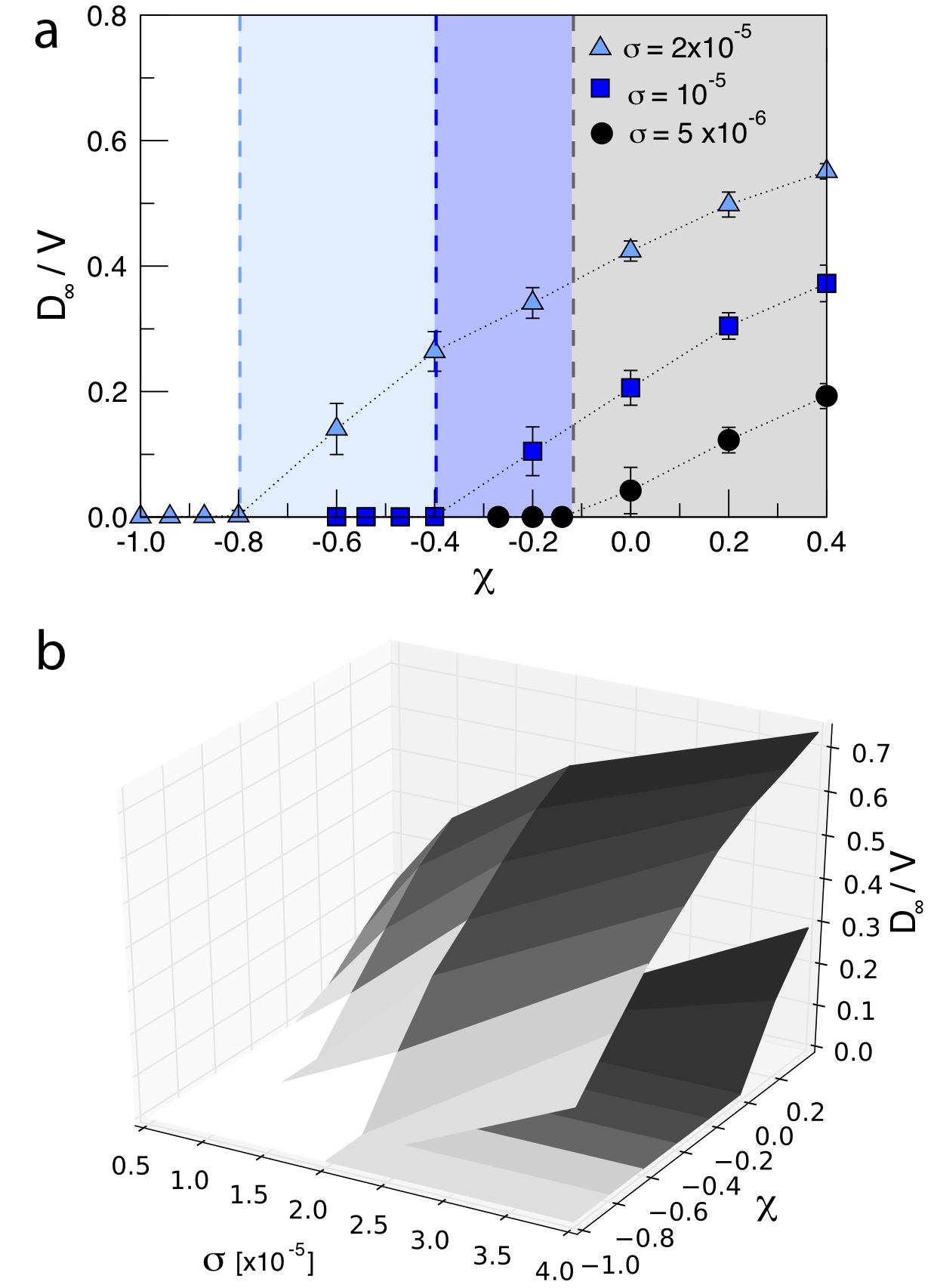}
\caption{\label{fig:th_num_het-hom} Global invasion transition as obtained by numerical simulations. Panel a: global attack rate, $D_\infty/V$, as a function of $\chi$ for different values of the diffusion rate rescaling $\sigma$. Colored areas highlight for each curve the invasion region defined by  Eq.~(\ref{eq:threshold}). Panel b: global attack rate, $D_\infty/V$, as a two dimensional function of $\chi$ and $\sigma$ for two different network topologies: a heterogeneous topology and a homogeneous one. Here $R_0=1.2$ and $\mu=0.002$.}
\end{center}
\end{figure*}

Here we present the results obtained with extensive Monte Carlo simulations in order to validate the theoretical analysis described above and test the robustness of our results under changes in the assumption on the timescale separation. 

We perform mechanistic numerical simulations at the individual level, keeping track of each individual's origin and destination and epidemic status over time. All transitions between compartments and the mobility events are modeled through discrete-time stochastic processes.
We create a network of $V$ subpopulations, where the connections between nodes are generated according to two different random graph topologies: a homogeneous and a heterogeneous one.
The homogeneous graphs~\citep{Erdos1959} are created by randomly wiring subpopulations with a constant probability $p = \langle k \rangle/(V-1)$, where $\langle k \rangle$ is the chosen average degree.
The heterogeneous graphs are scale-free networks, characterized by a power law degree distribution $P(k) \sim k^{-\gamma}$ and generated by the uncorrelated configuration model~\citep{Molloy1995,Catanzaro2005}, with $\gamma=3.0$ or $\gamma=2.1$ and $k_{\min}=2$.
Once the network topology is defined, the demographic quantities and the traffic patterns of the system are also defined. 
We assign to each node of degree $k$ a population $N \sim k^\phi$, with $\phi = \frac 3 4$ and we set the average population to $\overline{N}=10^3$. 
The leaving rate $\sigma_{kk'}$ is assumed to be proportional to $\sigma k^{\theta-\phi} k'^\theta$, as indicated in Eq.~(\ref{eq:diffusion_rate}), with $\theta = \frac 1 2$. 
Without changing the leaving rate distribution, we explore different length of stay distributions $\tau_k$ by varying the exponent $\chi$ in the range $[-1, 0.4]$ and keeping the average length of stay constant, $\overline{\tau}=37$ time steps. 
Values of $\chi$ and $\overline{\tau}$ are chosen in order to ensure the physical condition $\min(\tau_k) > 1$ and maintain feasible computational times. 
Moreover, in our simulations we explore different values of the average infectious period $\mu$ in order to test the limits of the timescale separation approximation.
Simulations are initialized with one randomly chosen subpopulation seeded by $I(0)=10$ individuals. We run $500$ realizations of the metapopulation model and follow the evolution of each outbreak until the disease dies out.
At the end of each realization we measure the global attack rate, $D_{\infty}/V$, that is the total fraction of infected subpopulation, which can be easily related to the global threshold condition. When the system is below the threshold, the global attack rate will fluctuate around the zero value, but, as the threshold condition $R_*>1$ is reached, the global attack rate will be significantly larger than zero.

Let us first focus on the relation between the global epidemic threshold $R_*$, the local epidemic threshold $R_0$, and the parameter $\chi$ that regulates the distribution of the length of stay.
In the numerical results, we recover the dependence of the critical value of the global threshold on $R_0$ and $\chi$, consistently with the analytical results of Fig.~\ref{fig:th_th}. 
In particular, as shown by the blue curve of Fig.~\ref{fig:th_num_het-hom}a, given the same network structure used in the theoretical analysis, if we assume $R_0=1.2$ and the diffusion rate rescaling $\sigma$ to be equal to $10^{-5}$ as in Fig.~\ref{fig:th_th}, the system reaches the invasion phase for a critical value of $\chi$ that well matches the theoretical value calculated from Eq.~(\ref{eq:threshold}) and indicated by the blue dashed line. 

In the same panel of Fig.~\ref{fig:th_num_het-hom}, we uncover the dependence of the global invasion threshold on the individual mobility rates, by exploring epidemic scenarios with different values of the diffusion rate rescaling $\sigma$. 
Consistently with the analytical results, for fixed values of $R_0$ and $\chi$, increasing values of the diffusion rate can bring the system into the invasion region. For the values of $R_0$ and $\sigma$ considered, when the length of stay is positively correlated with the degree of a subpopulation the system is always above the threshold, as predicted by the theoretical analysis (see Fig.~\ref{fig:th_th_het-hom_par}a). 
However, the final epidemic size is smaller for smaller values of $\sigma$, indicating that even if it is not possible to halt the spreading by lowering the mobility rates of individuals, it is possible to reduce the attack rate of the epidemic.
In order to highlight the good agreement between the individual-level simulations and the corresponding analytical results, for each curve we indicate the invasion region with a colored area, which is delimited on the left by a dashed line corresponding to the theoretical threshold value of $\chi$.

The network topology, as we have stressed before, plays a fundamental role in the disease spreading dynamics. In order to check the effect of the topology on the global invasion threshold, in the bottom panel of Fig.~\ref{fig:th_num_het-hom} we compare the numerical results obtained on a heterogeneous topology and a homogeneous one. In agreement with the analytical picture, in a homogeneous network the containment phase is achieved for a large range of  $\chi$ and $\sigma$ values, while a heterogeneous topology considerably favors the global invasion, leading to a non-zero global attack rate for almost all values of $\chi$ and $\sigma$ explored. 

It is important to stress that all the results presented so far have been derived under the assumption that $R_0-1 \ll 1$ and that the timescale associated to the disease is much longer than the mobility timescale, i.e. $\mu^{-1} \gg \tau_k$ for every degree $k$. 
The latter assumption would correspond, for instance, to consider the case of a typical influenza-like illness spreading on a network of locations visited during a typical human daily routine. 
The mobility timescale would be of the order of minutes or hours, while the disease timescale would be of a few days.
However, we want to test whether our theoretical framework can be applied also to the case of a network of cities connected by air travel, as it is often used for the study of the global spread of infectious diseases~\citep{Rvachev1985, Colizza2007a,Colizza2007d,Balcan2009b, Chao2010}, where the typical timescales of the mobility process and of the disease are comparable. In such case, the relation $\mu^{-1} \gg \tau_k$ would  no longer be valid, therefore we explore in the following the validity of our results when this assumption breaks down. 

\begin{figure*}[t]
\begin{center}
\includegraphics[width=9cm]{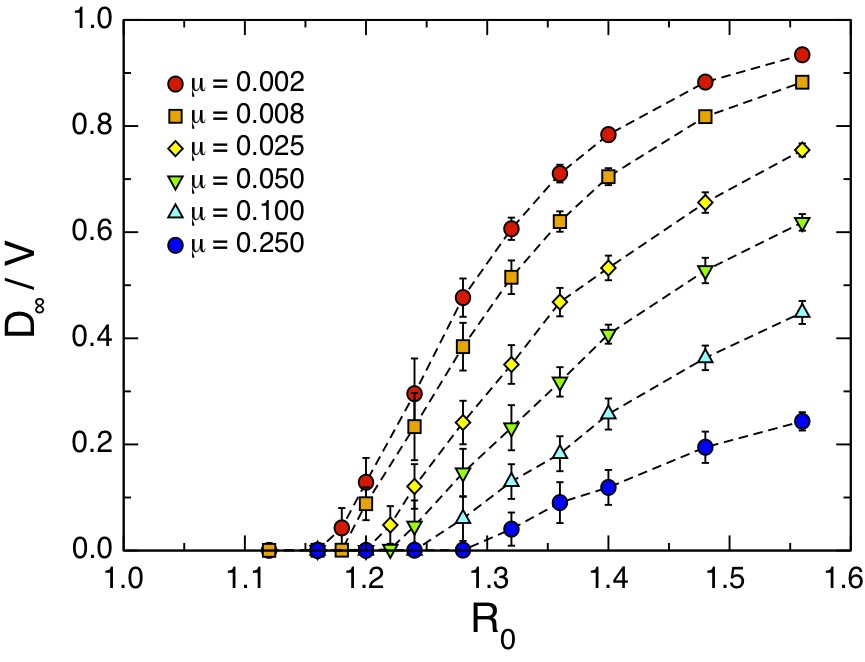} 
\caption{\label{fig:th_num_par} Global invasion transition as obtained by numerical simulations. Global attack rate as a function of $R_0$ for different values of $\mu$ with $\chi=-1$ and $\sigma= 4 \times 10^{-5}$.}
\end{center}
\end{figure*}

We run simulations for increasing values of $\mu$ in the range $[2\times10^{-3} - 2.5\times10^{-1}]$, by keeping the length of stay distribution constant with $\chi=1$ and $\overline{\tau}=37$.
As shown in Fig.~\ref{fig:th_num_par}, even in the case of a short infectious period like $\mu^{-1}=4$, we still observe a global phase transition as a function of $R_0$, indicating that our  description of the metapopulation system, characterized by two thresholds, is still valid. While the analytical result of Eq.~(\ref{eq:threshold}) did not predict any dependence of $R_*$ on $\mu$, the global threshold  in Fig.~\ref{fig:th_num_par} is  reached for different values of $R_0$ as $\mu$ varies, which is clearly due to the breakdown of the timescale separation approximation. The effect on the critical value of $R_0$ is however limited and corresponds to an approximate variation of $10\%$, even under a change of two orders of magnitude in $\mu$.

\section{Change of travel behavior due to infection\label{sec:oneseed}}
As discussed in Section \ref{sec:metapop}, the modeling framework we presented so far has been based on the assumption that individuals will always travel irrespective of their infectious status and associated symptoms.
However, this assumption is not completely realistic, since it is quite reasonable to assume that people would change their travel habits when ill. Modeling human reactions to the spread of infectious disease has recently attracted substantial attention~\citep{Funk2010}, also specifically with the inclusion of self-initiated behavioral changes into the mobility patterns of individuals in a spatially structured system~\citep{Meloni2011}.
Here, we study the effects of changes in the travel behavior of sick individuals on the condition for the global invasion that we derived in Section \ref{sec:global_th}.
In particular, we change our initial assumption by considering that individuals do not leave home if ill, but they can return home if they were infected at their destination during a trip.

The new assumption on individuals' behavior directly affects the expression of the global invasion threshold, Eq.~(\ref{eq:threshold}), through a change in the seeding mechanism between subpopulations.
In particular, given a diseased subpopulation of degree $k$ connected to a disease-free subpopulation of degree $k'$, the number of possible seeders is now represented only by susceptible individuals of $k'$ traveling to $k$, catching the disease and coming back infected.
Therefore, Eq.~(\ref{eq:2seeds}) changes to:
\begin{equation}
\lambda_{k'k} = \alpha \overline{N}_{k'k} \,,
\label{eq:1seed}
\end{equation}
where we still apply the quasi-equilibrium approximation.
Replacing the new expression of $\lambda_{k'k}$ in Eq.~(\ref{eq:branching}) we find that the global invasion threshold equation reduces to:
\begin{equation}
R_* = \frac{2(R_0-1)^2}{R^2_0} \sigma \overline{N} \overline{\tau} \frac{ \langle (k-1)k^{2\theta+\chi+1} \nu_k \rangle}{\langle k \rangle \langle k^\phi \rangle \langle k^\chi \rangle}\,,
\label{eq:th_1seed}
\end{equation}
where $\nu_k$ is still described by Eq.~(\ref{eq:pop2-deg}).

\begin{figure*}[tp]
\begin{center}
\includegraphics[width=\textwidth]{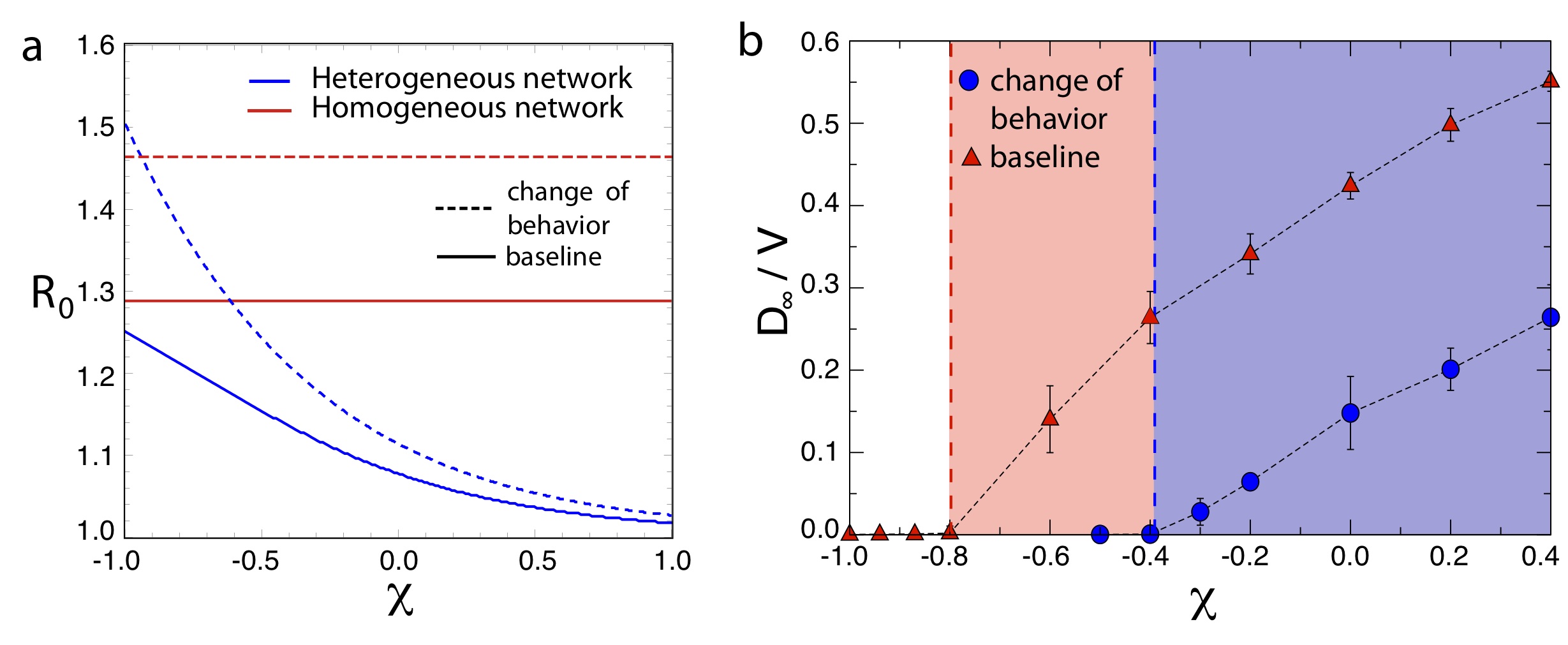} 
\caption{\label{fig:1seed_th2}  Effects of changes of travel behavior on the global epidemic threshold. Panel a: phase diagram defined by the threshold condition $R_*(R_0, \chi)=1$, for two mobility modeling assumptions: the baseline case corresponding to Eq.~(\ref{eq:threshold}) (solid line) and the behavioral change corresponding to Eq.~(\ref{eq:th_1seed}) (dashed line). Both models are run on a heterogeneous (blue lines) and homogeneous topology (red lines). Panel b: global attack rate, $D_\infty/V$, as a function of $\chi$ for two mobility modeling assumptions. Vertical dashed lines indicate the critical threshold value, as calculated from the curves shown in the left panel. The colored areas highlight the invasion region for each curve. The substrate network is scale-free with $\gamma=3$. Other parameters are: $R_0=1.2$, $\mu=0.002$, $\sigma=2 \times 10^{-5}$.} 
\end{center}
\end{figure*}

In Fig.~\ref{fig:1seed_th2}a we compare the phase diagrams of two metapopulation systems characterized by the same mobility parameters considered in Fig.~\ref{fig:th_num_het-hom} but two different seeding mechanisms: the baseline seeding framework, described by Eq.~(\ref{eq:2seeds}), and the above mentioned change of travel behavior described by Eq.~(\ref{eq:1seed}).
The curves represent the global invasion threshold condition $R_*=1$ in the $(R_0,\, \chi)$ parameter space. The invasion region is located above each curve, and the containment region below it.
As expected, the self-imposed limitations of individual mobility result in higher global thresholds with respect to the baseline case, both considering a heterogeneous topology and a homogeneous one. 
Intuitively, by reducing the number of possible seeding events, the disease invasion can only take place for higher transmissibility values. However, in a heterogeneous network this effect can be enhanced or suppressed depending on the assumed distribution of the length of stay. 
For positive and increasing values of $\chi$, travelers spend a significant amount of time in well connected nodes, which are also the main locations of transmission. In this case, preventing sick individuals from leaving home does not stop the progression of the disease through the hubs.
On the other hand, when the length of stay is negatively correlated with the connectivity of a node, i.e. $\chi<0$,  the longer time spent by travelers in poorly connected locations further reduces the number of transmission events. A change in the travel behavior of infectious individuals may then result in the containment of the disease, which could not be achieved in the corresponding baseline traveling scenario.

We validate the above theoretical results on the global epidemic threshold by means of Monte Carlo mechanistic simulations, following the methodology described in Section \ref{sec:global_th}. 
In the right panel of Fig.~\ref{fig:1seed_th2}, we compare the global attack rate $D_{\infty}/V$ as a function of $\chi$ obtained in the baseline scenario and shown in Fig.~\ref{fig:th_num_het-hom}a ($\sigma = 2\times 10^{-5}$) with the global attack rate measured in a scenario where sick individuals do not leave their homes.
Both scenarios are based on the same mobility parameters and differ only by the seeding mechanism under consideration.
The analytical results of the left panel are confirmed by the numerical simulations. The good agreement is highlighted by the color-coded areas that indicate the invasion regions. Dashed vertical lines indicate the critical values of $\chi$ calculated from Eq.~(\ref{eq:threshold}) and Eq.~(\ref{eq:th_1seed}), respectively.
The reduced mobility of sick individuals leads to the containment of the disease for a larger range of $\chi$ values, confirming the effects of such changes on the epidemic threshold.
Overall, it is evident that self-imposed behavioral changes related to the disease progression can have a significant impact on the epidemic spread in a spatially structured metapopulation model with memory, confirming similar results obtained within different mathematical frameworks~\citep{Perra2011,Funk2009,Scoglio2012,Poletti2009}.

\section{Epidemic spreading simulations above the invasion threshold \label{sec:above_th}}

\begin{figure*}[tp]
\begin{center}
\includegraphics[width=9cm]{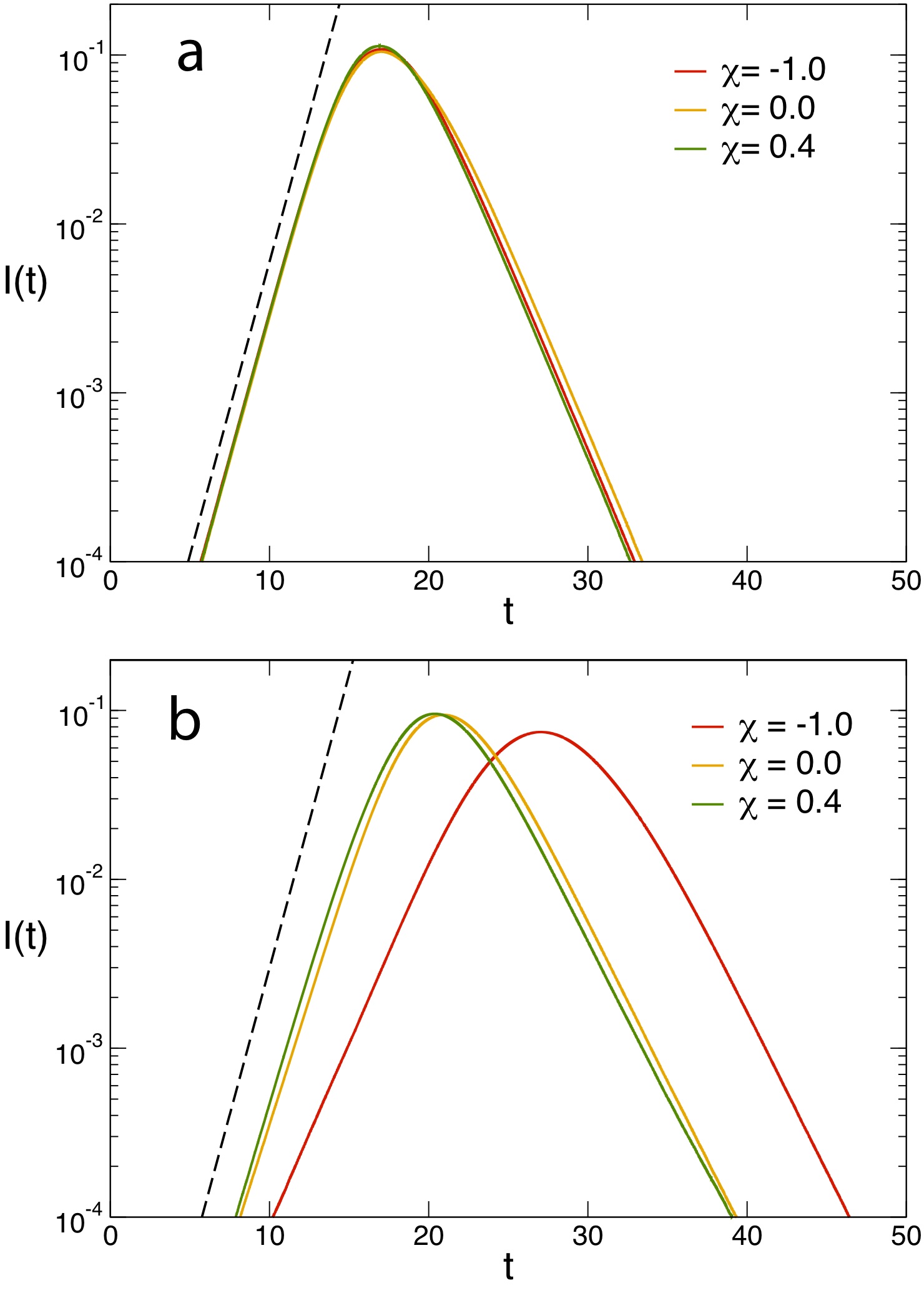} 
\caption{ Evolution of the fraction of infectious individuals for different values of $\chi$ and different mobility regimes. The time is measured in units of the average infectious period $\mu^{-1}$. When the leaving rate is high ($\sigma= 7\times 10^{-3}$, panel a) the parameter $\chi$ does not affect the exponential growth which is given by $\mu(R_0-1)$ (dashed line). For smaller values of $\sigma$  ($\sigma= 10^{-4}$, panel b), the role of $\chi$ becomes important. In both the cases $R_0=1.8$, the network is scale free with $\gamma=2.1$, $\mu= 0.002$, $\bar \tau= 37$.}
\label{fig:at_dynam}
\end{center}
\end{figure*}

In order to study the effects of the heterogeneous distribution of the mobility timescales on the epidemic dynamics, in addition to the invasion condition, we numerically study  the behavior of the metapopulation system above the invasion threshold, i.e. for $R_*>1$.
In this regime, the threshold condition assures that during the course of an outbreak a macroscopic fraction of subpopulations will be affected by the epidemic.  The spreading patterns may however be very different depending on the assumed distribution of the length of stay and its interplay with the network topology and mobility patterns.

We first focus on the time evolution of the global prevalence of the disease, $I(t)$, that is the total fraction of infectious individuals that are present in the system at a given time $t$.
It is possible to show, as detailed in \ref{app:global_prevalence}, that this quantity follows the equation:
\begin{equation}
\frac{dI(t)}{dt} = \beta \sum_{i \in V} S^*_i(t)\frac{I^*_i(t)}{N^*_i(t)} - \mu I(t) \,,
\label{eq:global_prevalence}
\end{equation}
where the sum runs over all the subpopulations and $S^*_i$ and $I^*_i$ represent all the susceptible and infectious individuals, respectively, who are present in a patch $i$ at time $t$; the population size of patch $i$ is indicated by $N^*_i$ and it accounts for all individuals who are located in $i$ at that timestep.

The spatial structure of the system is included in Eq.~(\ref{eq:global_prevalence}) through the non-linear terms $S^*_i(t) I^*_i(t)/N^*_i(t)$ that depend on the people locally present in node $i$ at each time step. 
In the case of a large number of coupled subpopulations, it is hard to derive an analytical solution of Eq.~(\ref{eq:global_prevalence}), however, we can simplify the picture by considering the infection dynamics in each subpopulation at the early stage, namely $S^*_i(t) \simeq N^*_i(t)$ for every $i$. 
We recover in this case the standard linear differential equation that describes the early-stage epidemic evolution in a single homogeneously mixed population, $d_tI(t) = (\beta - \mu)I(t)$.
The approximation of early stage infection evolution in all the subpopulations corresponds to the case in which the infection is rapidly seeded in all the subpopulations and the epidemic evolves synchronously across the system. Such description can be valid only if the mobility coupling between nodes is strong enough to allow a fast disease spreading through the network.
In this regime, the subpopulations are highly mixed and the spatial structure plays a minor role in the epidemic process; accordingly, we can expect that the distribution of mobility timescales does not significantly affect the spreading dynamics.

Numerical simulations confirm the above theoretical picture, as shown in Fig.~\ref{fig:at_dynam}. The plot displays the time behavior of the global prevalence $I(t)$ in a heterogeneous network of subpopulations and for different exponents $\chi$ of the length of stay distribution. 
In panel a, simulations consider a relatively high diffusion rate rescaling~--~$\sigma = 7 \times 10^{-3}$. Regardless of the values of $\chi$~--~positive, null or negative~--~the three curves follow exactly the same dynamics and their exponential growth is characterized by the coefficient $(\beta - \mu)$, indicating that for high mobility rates the global epidemic process can be effectively described by a single-population spreading dynamics. 
When the coupling among subpopulations is smaller, the strong coupling approximation is no longer valid. The spatial structure emerges and the effects of the network topology and traveling timescales on the epidemic dynamics are significant. Panel b of Fig.~\ref{fig:at_dynam} shows the effects of a change of almost two orders of magnitude  in the leaving rate rescaling, assuming $\sigma = 10^{-4}$ and keeping all the other parameters the same. 
Given the value of the basic reproduction number, $R_0=1.8$, the epidemic is still able to invade the entire system, but the dynamics is slower and, interestingly, completely different behaviors emerge according to different values of $\chi$: low negative values of $\chi$ significantly slow down the global invasion with respect to the case of null or positive values. 
This result reflects the impact of $\chi$ on the global threshold: when $\chi$ is positive, travelers spend more time in well frequented locations, which are also well connected, therefore facilitating the disease transmission among subpopulations. 
For negative values of $\chi$, travelers spend more time in less popular locations, reducing the probability of transmission at the global scale.

\begin{figure*}[tp]
\begin{center}
\includegraphics[width=9cm]{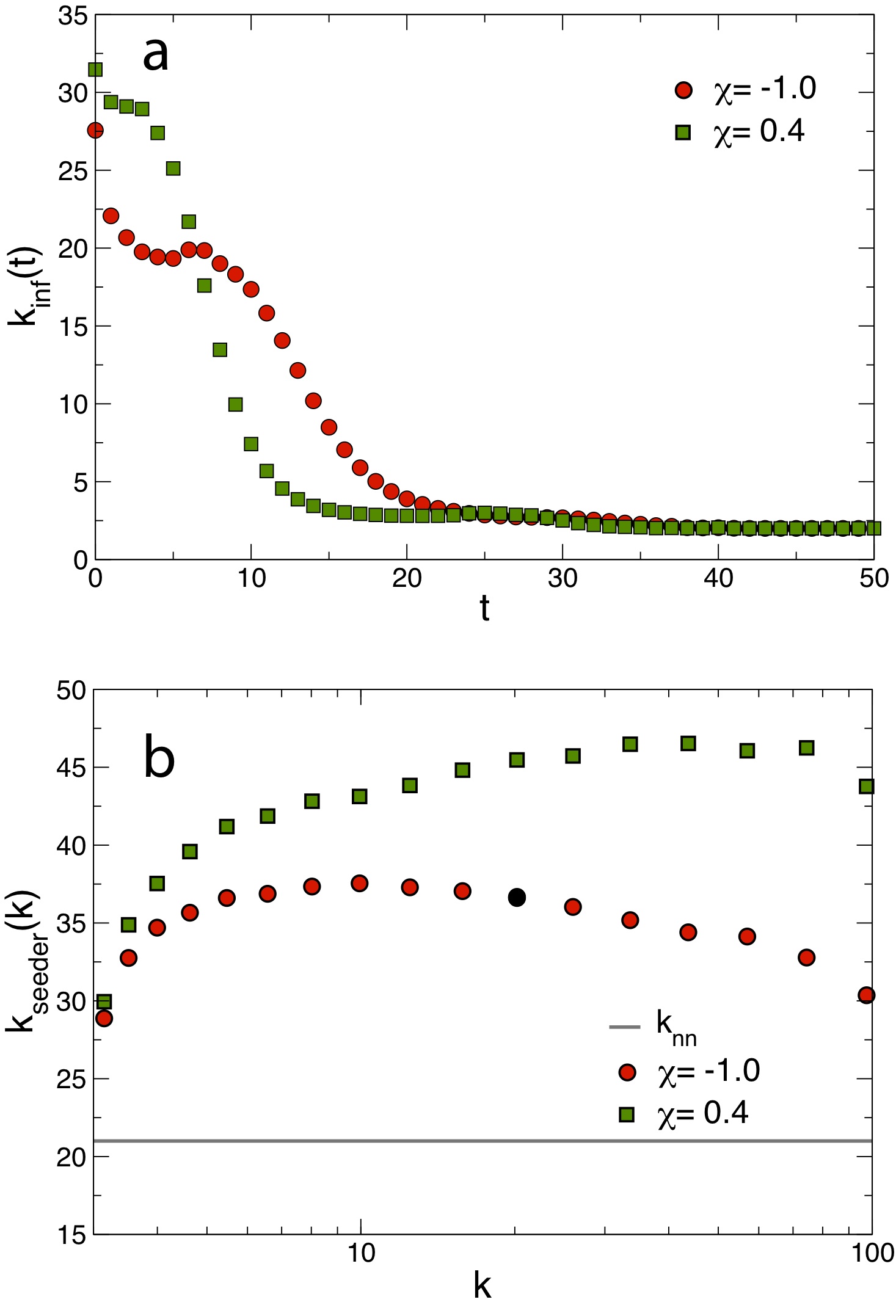} 
\caption{Panel a: average degree of newly infected subpopulations as a function of time,  $k_{\textmd {\scriptsize{inf}}}(t)$. The quantity decreases as the epidemic invades the system, which indicates that the epidemic diffuses from the high-degree nodes to the small-degree ones. Positive values of $\chi$ enhance this effect. Panel b: average degree of the seeder node as function of the seeded node degree, $k_{\textmd {\scriptsize{seeder}}}(k)$. By increasing the value of $\chi$ the contribution of the high-degree nodes to the spreading process increases. The grey line indicates the average nearest neighbor degree $k_{\textmd {\scriptsize{nn}}}$, as a reference. Here $R_0= 1.8$ and $\sigma= 10^{-4}$.  The substrate network is scale-free with $\gamma= 2.1$. All the other parameters are the same as in Fig.~\ref{fig:at_dynam}.}
\label{fig:at_cascade}
\end{center}
\end{figure*}

When the spatial component becomes relevant, that is for small values of the diffusion rates, the effects of the heterogeneous distribution of mobility timescales can be observed in the invasion pattern from one infected subpopulation to the others.
In our numerical simulations, we focus on the average degree of the newly infected nodes at each time step, $k_{\textmd {\scriptsize{inf}}}(t)$, which is defined as:
\begin{equation}
k_{\textmd {\scriptsize{inf}}}(t)=\frac{\sum_k [D_{k}(t) - D_{k}(t-1)]}{D(t) - D(t-1)} \,,
\label{eq:kinf}
\end{equation}
where $D_{k}(t)$ represents the number of subpopulations with degree $k$ that have been infected at time $t$. 
As already pointed out in~\citep{Barthelemy2005}, this observable provides a good description of the spreading pattern, highlighting the role of the network topology in shaping the epidemic propagation. 
Scale-free networks are characterized by the so called cascade effect~\citep{Barthelemy2005}: the highest-degree nodes are quickly reached by the infection and only afterwards the epidemic reaches the nodes with smaller degree. 
In the top panel of Fig.~\ref{fig:at_cascade}, we provide evidence for this behavior in the case of a non-Markovian dynamics on a heterogeneous metapopulation network. 
The quantity $k_{\textmd {\scriptsize{inf}}}(t)$, as measured in the numerical simulations, is plotted for two distinct values of $\chi$, $\chi=0.4$ and $\chi=-1$. 
The cascade effect, due to the scale-free topology of the substrate network, is evident in both cases: at the early stage of the epidemic the average degree of newly infected nodes is high, but, as the spreading process evolves in time, $k_{\textmd {\scriptsize{inf}}}(t)$ decreases towards the smallest degree value, $k_{\min}=2$. 
Even if qualitatively similar, the two curves corresponding to different distributions of the length of stay show some relevant differences: for $\chi=0.4$ the epidemic rapidly reaches all the nodes with the largest degree and, after this early stage, the curve has a sharp drop down to values close to $k_{\min}$. 
In the case $\chi=-1$, such drop is less pronounced, indicating that a short length of stay in high degree nodes reduces their probability to be seeded by the infection. In this situation the role of the hubs as super-spreaders is reduced, leading to a global delay in the epidemic propagation.

In order to further explore the effects of the length of stay on the invasion patterns, we study the changes in the spreading potential of high degree nodes related to the variations of the length of stay parameter, $\chi$. 
In our numerical simulations, we track the degree of the infection seeder for each infected node $i$, $k_{\textmd {\scriptsize{seeder}}}$, that is the degree of the nearest neighbor subpopulation which is the source of the infection of $i$. 
This quantity is plotted in the bottom panel of Fig.~\ref{fig:at_cascade} as a function of the degree of the seeded node, $k$, averaged over each degree block. 
Since the network is uncorrelated, $k_{\textmd {\scriptsize{seeder}}}(k)$ should not depend on $k$, however, we observe that the functional dependence of the mobility parameters on the degree introduces non-trivial correlations between nodes, and eventually, the curve  $k_{\textmd {\scriptsize{seeder}}}(k)$ is not flat.
Overall, the value of $k_{\textmd {\scriptsize{seeder}}}(k)$ is very high compared to the average nearest neighbor degree $k_{\textmd {\scriptsize{nn}}}$, because the hubs are responsible for multiple seeding processes: the entire system is infected by a few number of high-degree nodes which are rapidly reached by the epidemic at the beginning and become the most important vehicle of infection spreading. 
This mechanism, which is typical of heterogeneous networks, is affected by the assumed distribution of the length of stay: by lowering the value of $\chi$, the average values of $k_{\textmd {\scriptsize{seeder}}}(k)$ are reduced, as the role of the hubs becomes less relevant.

\begin{figure*}[tp]
\begin{center}
\includegraphics[width=\textwidth]{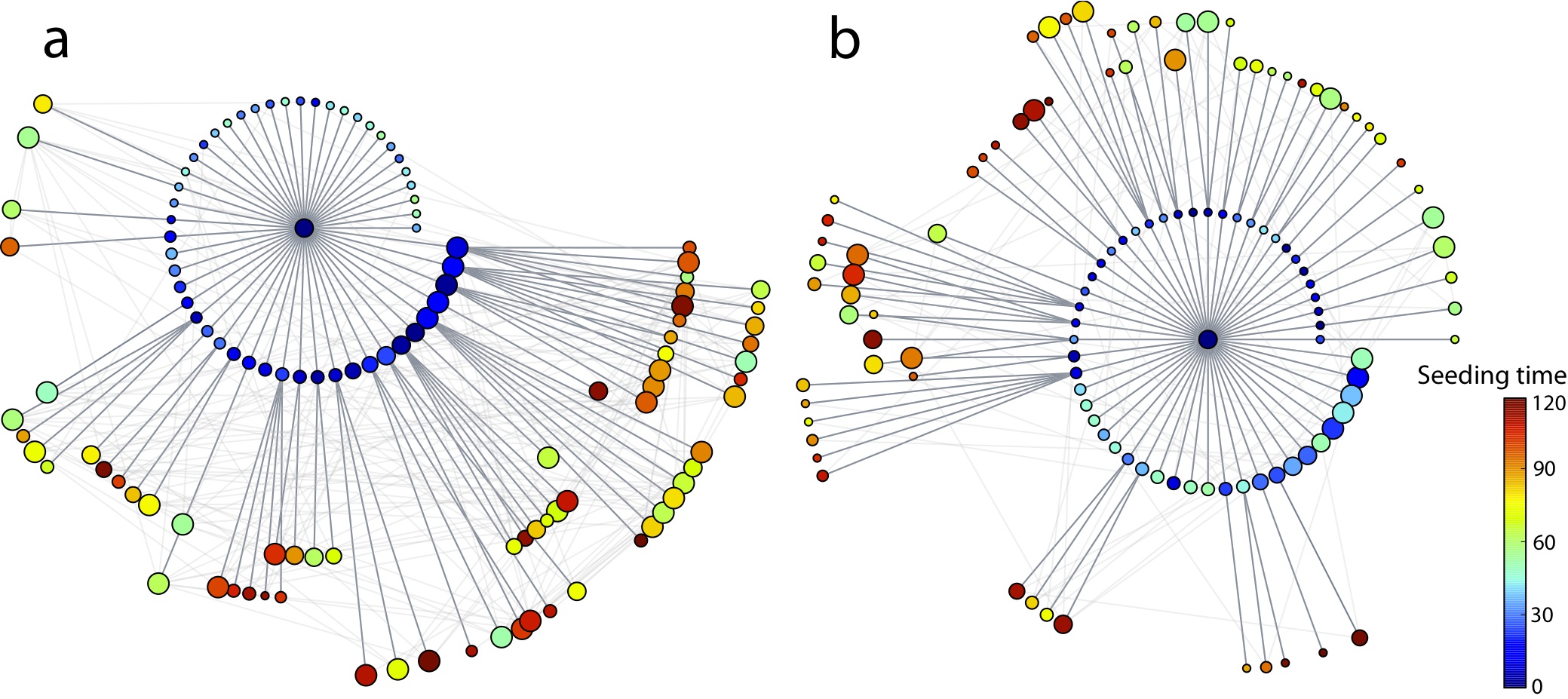} 
\caption{Epidemic invasion trees. The cases of a positively-correlated ($\chi=0.4$, panel a) and negatively-correlated ($\chi=-1$, panel b) length of stay are shown. The synthetic network is characterized by a power-law distribution with $\gamma= 2.1$. An SIR dynamics starting from the same seeding node (at the centre of each visualization) is simulated, with $R_0=1.8$, $\sigma= 10^{-4}$,  $\mu= 0.002$, $\bar \tau= 37$ to ensure the timescale approximation. Only the first 120 nodes to be infected are displayed for the sake of visualization, on successive layers of invasion. Larger width grey links correspond to the paths of infection and lighter grey ones to the existing connections among visible nodes. Nodes are color coded according to the time of their seeding, and their size scales with their degree; nodes in the first layer are ordered according to their degree to highlight the role of different degree nodes in the hierarchical invasion pattern in the two cases. }
\label{fig:at_sp} 
\end{center}
\end{figure*}

In order to summarize in a single picture all the differences in the sequences of transmission events obtained with different distributions of the length of stay, we also show a visual representation of the seeding process on a heterogeneous network.
In this analysis, we compare simulated epidemics starting from the same initial conditions, randomly choosing a node to be seeded, and using different values of $\chi$. 
For each simulation, we build the epidemic invasion tree that represents the most probable transmission path of the infection from one subpopulation to the other during the course of the outbreak~\citep{Balcan2009b}. 
For each subpopulation pair $lj$, we define $p_{lj}$ as the probability of infection transmission from $l$ to $j$. This probability shows the likelihood that the infection in $j$ is seeded by subpopulation $l$. 
This can happen by two means: either a resident in $j$ acquires the infection in $l$ and brings it home, or an infectious traveller from $l$ brings the infection to $j$. Then, $p_{lj}$ is defined as the proportion of runs where $j$ has been seeded by $l$. 
Finally, we define a distance metric $d_{ij}=\sqrt{(1-p_{lj})}$ to measure dissimilarities for the infection probability, and we extract the directed weighted minimum spanning tree using the Chu-Liu-Edmunds Algorithm~\citep{Chu1965}, in order to eliminate loops and highlight the main directed paths of transmission.

Fig.~\ref{fig:at_sp} displays two epidemic invasion trees, extracted from a set of 100 simulations, and two different values of $\chi$: $\chi=-1$ and $\chi=0.4$
The origin of the infection is located at the centre of the tree, with successive generations of infected nodes mapped out as circular layers.
In the first layer, the nodes are ordered by degree (size of the dot) and by seeding time (color), showing how different values of $\chi$ alter the hierarchy of the epidemic invasion. 
For $\chi>0$, largely connected subpopulations are infected first and have a predominant role in the further spatial spread of the disease, thanks to the two-fold favouring property of having a high degree and a longer visiting time. 
On the other hand, for $\chi<0$,  the spreading potential of the hubs is suppressed by the very short length of stay of visitors. Less connected subpopulations become instead mainly responsible of the spreading dynamics towards the rest of the system.

\section{Markovian vs. non-Markovian mobility model \label{sec:mark_nomark}}
In order to further highlight the impact of the heterogeneous distribution of mobility timescales on the epidemic spread, we compare the results of our analytical and numerical non-Markovian framework with a standard Markovian approach.
The behavior of Markovian reaction-diffusion processes on a metapopulation system connected by a heterogeneous network has been extensively investigated~\citep{Colizza2007b, Colizza2007c, Colizza2008}. 
In general, we can expect to observe a faster disease spread and a wider spatial invasion, at a given time step, in memoryless models, because they imply a higher degree of mixing between individuals~\citep{Keeling2010, Balcan2012}.

Here, we want to quantify the differences between the two approaches by comparing the same metapopulation system with two mobility dynamics and focus on (i) the change in the global invasion threshold as a function of the local threshold, $R_0$, and (ii) the change in the spreading pattern between subpopulations. 
To this aim, we create a metapopulation model with a non-Markovian mobility dynamics as described in Section \ref{sec:metapop} and  compare it to an identical metapopulation system having the same traffic volumes along each connection, but  ruled by Markovian mobility equations.
Once the leaving rate and the length of stay in the non-Markovian model are defined, it is possible to compute $T_{kk'}$, the total volume of people traveling along each link at the equilibrium, that is the sum of people resident in $k$ and travelling to their destination $k'$, and people resident in $k'$ returning after visiting $k$:
\begin{equation}
T_{kk'}= \sigma_{kk'} N_{kk} + \tau_{k}^{-1} N_{k'k}
\end{equation}
By inserting the expressions of Eq.~(\ref{eq:pop0-deg}) and Eq.~(\ref{eq:pop1-deg}), we find:
\begin{equation}
T_{kk'} = \frac{\sigma \overline{N}}{\langle k^\phi \rangle} (kk')^\theta (\nu_{k} + \nu_{k'})\,.
\label{eq:traffic_volume}
\end{equation}
We can therefore construct a Markovian mobility model in which the traffic through each link is defined by Eq.~(\ref{eq:traffic_volume}) but no distinction is made between the individuals that are traveling to a subpopulation $i$ and the individuals resident in $i$: all individuals regardless their origin have equal probability to leave node $i$ for each of the $k_i$ neighboring destinations.

\subsection{Global invasion threshold}
\begin{figure*}[tp]
\begin{center}
\includegraphics[width=9cm]{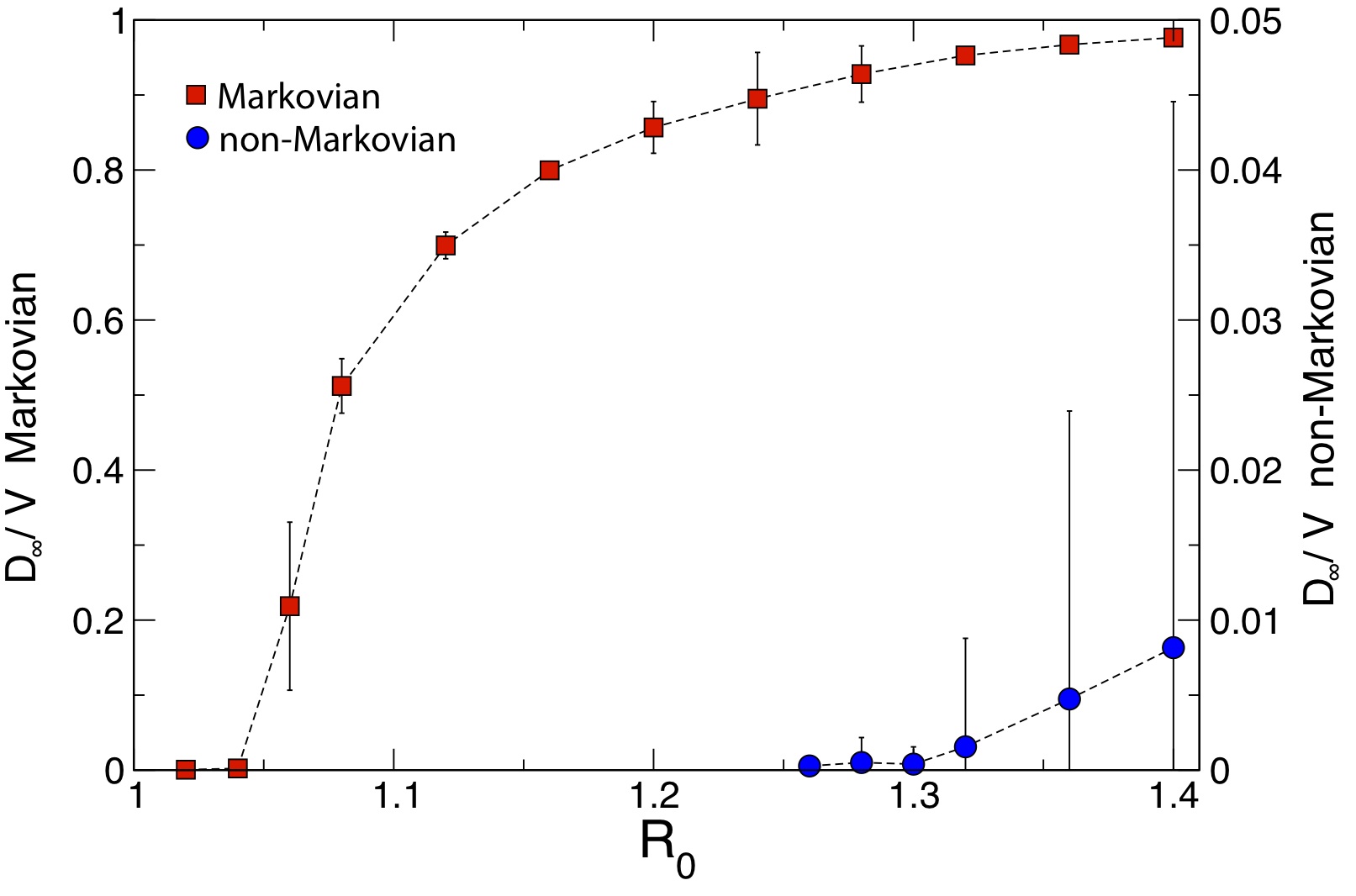} 
\caption{\label{fig:th_num_m-nm} Impact of the mobility model on the threshold behavior: comparison between the Markovian and the non-Markovian mobility dynamics. Fraction of infected cities $D_\infty/V$ as a function $R_0$ on a heterogeneous network with power-law degree distribution $P(k) \propto k^{-\gamma}$ and $\gamma=3$. The curve corresponding to a non-Markovian model with $\chi = -1$ is compared with the case of a Markovian mobility dynamics, where the traffic volume along each link is equal in both cases. The diffusion rate rescaling is $\sigma=10^{-5}$ and the average infectious period is $\mu=0.002$.}
\end{center}
\end{figure*}

We first focus on the behavior of the global attack rate, $D_{\infty}/V$, in the two modeling approaches as a function of the local epidemic threshold $R_0$.
In Fig.~\ref{fig:th_num_m-nm} we compare the global attack rate measured on a heterogeneous network and using a non-Markovian dynamics to the curve obtained with a Markovian traveling dynamics characterized by the same mobility parameters on the same network structure. 
The difference between the two mobility models is striking: in the Markovian dynamics case, the fraction of infected subpopulations rapidly increases reaching the whole network for $R_0=1.4$. 
On the other hand, in the non-Markovian case the epidemic is contained for almost all the explored values of $R_0$ and reaches a very limited fraction of subpopulations (at most $4\%$) for the same value of $R_0$.
As expected, large fluctuations characterize the measured attack rates when the system is close to the epidemic threshold in both cases. 
These results highlight the strong difference between the two models that are characterized by the same pattern of mobility flows but differ in the mobility mode of each single individual.
A global epidemic threshold is observed in both cases, but the absence of memory in the model significantly lowers the condition on $R_0$ corresponding to the containment phase.  
These findings have a significant impact on the modeling of infectious disease in general, since it is evident that, when a Markovian approach is adopted, the final impact of an epidemic will be potentially overestimated if the role of mobility timescales is ignored.

\subsection{Epidemic dynamics above the threshold}
We further explore the difference between non-Markovian and Markovian modeling approaches by looking at the invasion regime above the global invasion threshold.
In particular, we compare the global prevalence curve obtained using our non-Markovian framework for the two cases $\chi = 0.4$ and $\chi=-1$, as displayed in Fig.~\ref{fig:at_dynam}b, with the results of a Markovian mobility dynamics, where the traffic volume along each link is kept the same as the non-Markovian case.
It worth to notice that, from Eq.~(\ref{eq:traffic_volume}), the traffic volume on a link depends on the exponent $\chi$, therefore also the average traffic of the system is a function of the exponent $\chi$.
The average traffic can be computed analytically by averaging  the expression given by Eq.~(\ref{eq:traffic_volume}) over all links of the system:
\begin{equation}
\langle T_{kk'} \rangle= \sum_{kk'} \frac{kP(k)k'P(k')}{\langle k \rangle^2} \frac{\sigma \overline{N}}{\langle k^\phi \rangle} (kk')^\theta (\nu_{k} + \nu_{k'})\,,
\end{equation}
where we assumed that the network is uncorrelated. By averaging over the degrees $k$ and $k'$, we obtain the final expression:
\begin{equation}
\langle T_{kk'} \rangle = \frac{2 \sigma \overline{N}}{\langle k \rangle ^2 \langle k^\phi \rangle} \langle k^{\theta+1} \nu_{k}(\chi) \rangle  \langle k^{\theta+1} \rangle \,,
\end{equation}
where we have highlighted the dependence on $\chi$.
$\langle T_{kk'} \rangle$ decreases as the value of $\chi$ increases with a relative change that depends on the value of $\sigma$.
Therefore, in the comparison between Markovian and non-Markovian dynamics we do not keep the total traffic per link constant in the Markovian case, but we change it according to the value of $\chi$ under study.

\begin{figure*}[tp]
\begin{center}
\includegraphics[width=9cm]{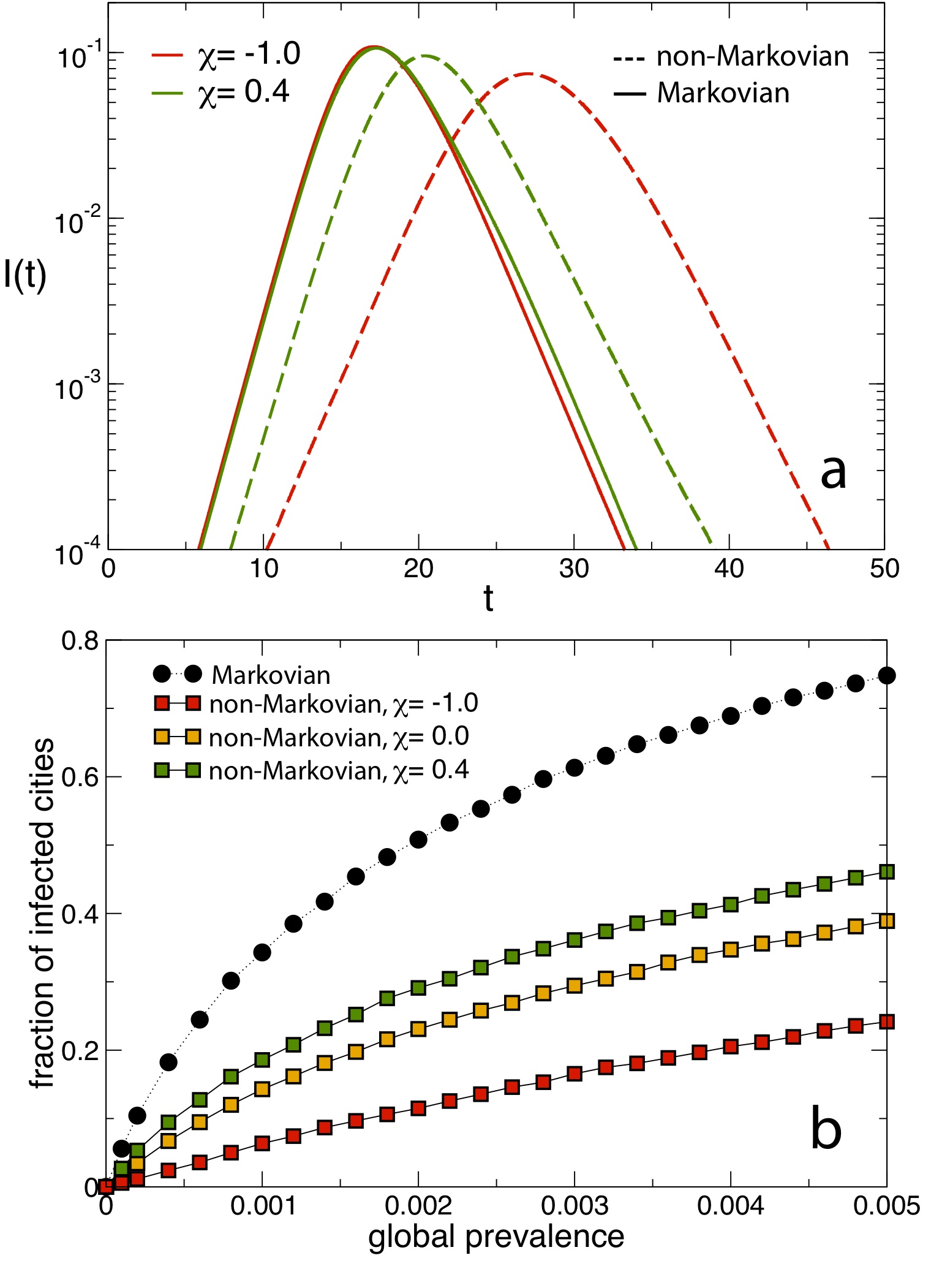} 
\caption{Impact of the mobility model on the epidemic spreading above the invasion threshold: comparison between the Markovian and the non-Markovian mobility dynamics. Panel a: the curves of Fig.~\ref{fig:at_dynam}b, corresponding to the case $\chi=-1$ and $\chi=0.4$ are compared with the results of a Markovian mobility dynamics, where the traffic volume along each link is kept the same. Dashed and continuous curves corresponds to the non-Markovian and Markovian case respectively. Panel b: fraction of infected cities as a function of the global prevalence for different mobility models and distributions of the length of stay. The Markovian mobility model shows always the largest number of infected cities for a given global prevalence. Model parameters are: $R_0= 1.8$, $\mu= 0.002$, $\sigma= 10^{-4}$, $\overline{\tau}= 37$.}
\label{fig:at_m-nm}
\end{center}
\end{figure*}

The comparison of the global prevalence curves in the two models, displayed in Fig.~\ref{fig:at_m-nm}a, points out the strong enhancement in the epidemic spreading potential resulting from a Markovian traveling dynamics: the infection curve grows exponentially as in a single homogeneously mixed population, while the epidemic growth is significantly delayed in a non-Markovian approach. 
Moreover, the above results highlight the different role played by the exponent $\chi$ in the two dynamics. When the traveling dynamics is Markovian no appreciable difference exists between the two values of $\chi$, as shown by the solid lines. 
The variation in the traffic volume along the links, corresponding to the change of $\chi$, is too small to produce a significant effect at the global level. 
On the other hand, in the non-Markovian case different values of the exponent $\chi$ correspond to very different epidemic patterns: a result that is not simply due to modifications of the traffic volumes but is a consequence of the interplay between the heterogeneous traveling timescales and the network topology. A simple Markovian mobility model is not able to capture this particular aspect because it does not take  into account the trip duration.

Furthermore, the Markovian dynamics is not only characterized by a faster progression of the disease invasion but also by a larger number of affected subpopulations, for a given value of the global prevalence. 
In other terms, even if the total number of infectious individuals at a given time step is the same, they will be distributed among a much larger number of subpopulations in the Markovian case than in the non-Markovian one.
The bottom panel of Fig.~\ref{fig:at_m-nm} clearly illustrates this phenomenon.
The fraction of infected subpopulations in the system is shown as a function of the global prevalence for four different cases: a Markovian mobility dynamics and a non-Markovian mobility dynamics with three different values of the exponent $\chi$.
The Markovian model is characterized by the highest geographic dispersal, which largely exceeds the values observed in a non-Markovian model with $\chi=0.4$. On the other extreme, for negative values of $\chi$ the disease spread is confined to a relatively small set of locations.

\begin{figure*}[tp]
\begin{center}
\includegraphics[width=\textwidth]{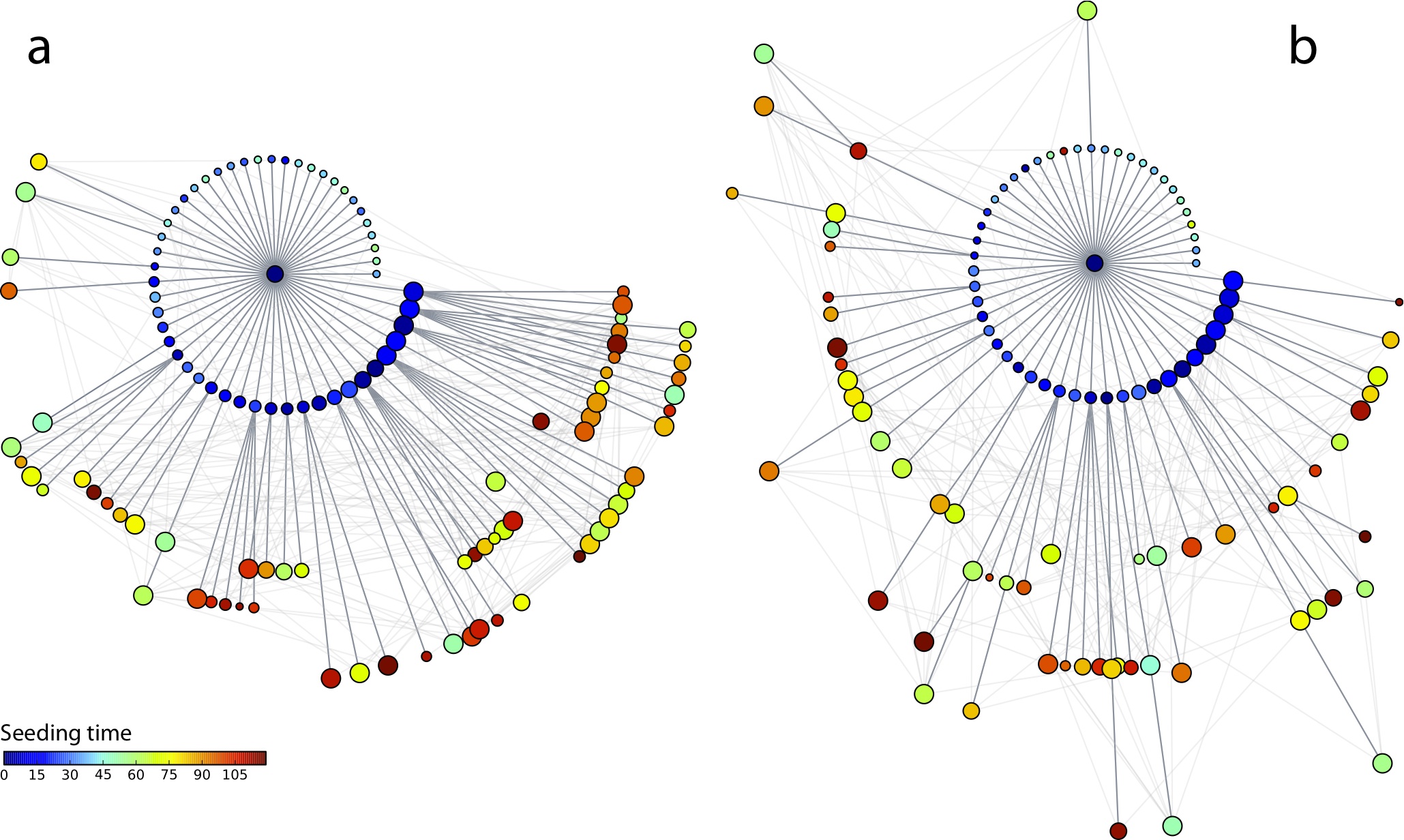} 
\caption{Epidemic invasion trees. The cases of a non-Markovian (panel a) and Markovian (panel b) mobility process for a given distribution of the length of stay ($\chi=0.4$) are shown. The parameters characterizing the substrate network and the dynamical process are the same of Fig.~\ref{fig:at_sp}. Only the first 120 nodes to be infected are displayed for the sake of visualization, on successive layers of invasion. Larger width grey links correspond to the paths of infection and lighter grey ones to the existing connections among visible nodes. Nodes are color coded according to the time of their seeding, and their size scales with their degree; nodes in the first layer are ordered according to their degree to highlight the role of different degree nodes in the hierarchical invasion pattern in the two cases. }
\label{fig:at_sp_m-nm} 
\end{center}
\end{figure*}


In order to examine a complete picture of the two mobility dynamics, we also compare the invasion trees obtained with a Markovian approach and a non-Markovian one. In particular, we compare the invasion tree of a non-Markovian spreading process with $\chi=0.4$ and shown in panel a of Fig.~\ref{fig:at_sp}, with the invasion tree of a spreading process that starts from the same node in the same network and same mobility parameters, but characterized by a Markovian dynamics (bottom panel of Fig.~\ref{fig:at_sp_m-nm}). 
Even if the hubs play an important role in both  spreading processes, there are some relevant differences that can be observed in the invasion trees. First, in the Markovian model the infection progresses from small degree nodes to high degree nodes with higher probability than in the non-Markovian case. 
This is due to the fact that the non-Markovian dynamics with positive $\chi$ strongly suppresses the spreading potential of small degree nodes. 
More importantly, the Markovian dynamics is characterized by a deeper extension of the tree, for a given number of infected nodes. Indeed, in the non-Markovian case the first $120$ infected nodes all belong to the first and second shell of tree, while in the Markovian case, the infection quickly reaches  the third shell. 
Given that the network has diameter equal to $10$, this result indicates that in the Markovian model the epidemic can more easily reach nodes that are far from the seed in terms of connections. 
Relevant differences are also evident in terms of time of infection, as indicated by nodes' color: in the non-Markovian case the nodes of the first shell are the first to be infected and only afterwards the infection progresses to those of the second shell. 
On the other hand, the Markovian dynamics accelerates the spreading process so that  nodes belonging to the third shell of the invasion tree can be infected before  nodes of the first shell, as indicated by the colors.

\section{Conclusions}
In this paper we presented a general theoretical framework to include a heterogeneous distribution of mobility timescales related to the length of stay at destination, as observed in reality, into a metapopulation epidemic model.
Our work is built upon the general theory of non-Markovian dynamical processes on metapopulation models, originally developed by Sattenspiel and Dietz~\citep{Sattenspiel1995}, and extends the mathematical framework of degree-block variables~\citep{Colizza2007b, Colizza2008} and the timescale separation approach~\citep{Keeling2002}.
While previous studies of non-Markovian mobility processes were based on the assumption that the mobility timescale is the same for every individual and across the whole system~\citep{Balcan2012}, here, prompted by empirical evidence, we introduced an extra layer of heterogeneity characterizing the host dynamics in terms of the duration of their visits~\cite{Poletto2012} and fully characterize the invasion condition for the metapopulation system and its dynamics above the threshold, exploring change of travel behaviors and differences with respect to Markovian approaches .

It is well known that large fluctuations in the number of connections and the number of travelers per connection have a strong impact on the spreading dynamics, in particular they favor the epidemic invasion by lowering the global epidemic threshold.
The integration of a non-Markovian dynamics with a heterogeneous distribution of mobility timescales reproduced similar results as those observed within a Markovian framework, but uncovered some relevant differences. 
We found that the metapopulation system is still characterized by a global threshold, above which a disease seeded in a single location can reach a finite fraction of subpopulations. 
However, given the assumed dependence of the length of stay with the subpopulation degree, both the theoretical framework and the applied numerical simulations have shown that two regimes are found that may dramatically favour or hinder the invasion, induced by the positive or negative degree-correlation of the length of stay, respectively, altering the predictions of simple Markovian models.
Moreover, we showed that the interplay between the connectivity of each node and the assumed distribution of mobility timescales has a profound impact on the epidemic invasion patterns when the system is above the threshold. 
In case of a length of stay that is negatively correlated with the degree, the spreading potential of the hubs is substantially reduced, leading to a strong delay of the epidemic invasion which also corresponds to a lower spatial dispersal of the disease. On the other hand, when travelers spend more time in highly connected locations the epidemic spread is accelerated by the dominant role of the hubs.
A further comparison with the classic Markovian dynamics highlighted the strong differences between the two approaches: in a memoryless model, due to the high degree of mixing between individuals and subpopulations, the epidemic progression occurs much faster and reaches a much larger fraction of subpopulations for a given prevalence at the system level. This difference might have a strong impact when interpreting the results of epidemic models.
Eventually, the non-Markovian theoretical framework allowed us to study the effects of self-imposed behavioral changes of individual mobility. We explored a simple scenario, where individuals do not leave home if sick but travel back to their residence if they were infected abroad. Despite its simplicity, this change of travel behavior affected the general results on the global epidemic threshold, confirming that the individual response to an epidemic outbreak is an important aspect that should be considered in simulating epidemic spreading patterns, and in providing detailed model predictions. 

Though applied here to human mobility and human epidemics, our approach is however valid for other hosts displaying a territorial nature linked to a permanent population, and with movements characterized by varying timescales. 
Our contribution presents  a number of limitations that need to be addressed in future work. In particular, the assumption on the geographical dependence of the length of stay, determined by the degree $k$ of the location, finds its support in travel statistics available at the city level but there is need for higher resolution mobility data to better characterize the length of stay distribution and include additional realistic aspects, such as e.g. the dependence of mobility rates on the age of travelers~\citep{Apolloni2013}, on the distance traveled, and others. For instance, more sophisticated assumptions can be made on the expression of $\tau$, that may depend both on the origin and destination subpopulations, or on the individual behavior.

As the spatial spread plays a crucial role in the management and control of a disease, our results highlight the importance of the mobility timescales in the epidemic dynamics and pave the way to the development of more realistic mathematical and computational epidemic models that could be used to support public health agencies in decision making.

\section*{Acknowledgements}
This work has been partially funded by the ERC Ideas contract n.ERC-2007-Stg204863 (EPIFOR) to V.C., C.P., and M.T; the EC-Health contract no. 278433 (PREDEMICS) to C.P. and V.C.; the ANR contract no. ANR-12-MONU-0018 (HARMSFLU) to V.C.

\appendix
\section*{Appendices}

\section{Populations at equilibrium\label{app:populations}}
Here, we provide some further details on the derivation in the degree-block notation of the equilibrium relations for the populations $\overline{N}_{kk}$ and $\overline{N}_{kk'}$ expressed by Eq. (\ref{eq:pop0-deg}) and Eq.( \ref{eq:pop1-deg}). 
We start from the equilibrium condition applied to Eq. (\ref{eq:no-mark1-deg}): 
\begin{equation}
\sigma_{kk'} N_{kk} -  \tau_{k'}^{-1} N_{kk'}=0,
\label{eq:no-mark2db}
\end{equation}
which yields:
\begin{equation}
N_{kk'}=\tau_{k'} \, \sigma_{kk'} N_{kk}.
\label{eq:Nkk'_from_Nkk}
\end{equation}
We now recall that the number of individuals resident in a subpopulation of degree $k$, $N_k$, can be expressed in the degree-block notation as:
\begin{equation}
N_k \equiv \frac {\bar N}{\langle k^\phi \rangle} k^\phi= N_{kk} + k \sum_{k'} \frac{k' P(k')}{\langle k \rangle} N_{kk'},
\label{eq:pop_db}
\end{equation}
where we assume that the network is uncorrelated, i.e. $P(k'|k) = k' P(k')/\langle k \rangle$.  
By plugging Eq.~(\ref{eq:Nkk'_from_Nkk}) into Eq.~(\ref{eq:pop_db}) and remembering the definitions of leaving rate and length of stay:
\begin{eqnarray}
\sigma_{kk'} &=& \frac{w_{kk'}}{N_k}= \sigma  k^{\theta-\phi} k'^\theta\\
\tau_{k'}&=&\frac{\bar \tau}{\langle k^\chi \rangle} k'^\chi.
\end{eqnarray}
we obtain the first equation for the stationary population $\overline{N}_{kk}$:
\begin{equation}
\frac {\bar N}{\langle k^\phi \rangle} k^\phi= \overline{N}_{kk} \left(1+ \frac{\sigma \bar \tau}{\langle k^\chi \rangle} \, k^{\theta-\phi+1} \sum_{k'} \frac{P(k')}{\langle k \rangle}  k'^{\theta+\chi+1} \right).
\label{eq:almost}
\end{equation}
The term within round brackets is inverse of the quantity $\nu_k$ as defined in Eq.~(\ref{eq:pop2-deg}), therefore Eq.~(\ref{eq:almost}) provides the expression of $\overline{N}_{kk}$ given in Eq.~(\ref{eq:pop0-deg}). 
The expression of $\overline{N}_{kk'}$, as given by Eq.~(\ref{eq:pop1-deg}), follows directly by plugging Eq. (\ref{eq:pop0-deg}) into the equilibrium relation Eq. (\ref{eq:Nkk'_from_Nkk}).


\section{Branching process and global invasion threshold \label{app:threshold}}
The number of diseased subpopulations of degree $k$ at generation $n$ can be related to those at generation $n-1$ by the equation:
\begin{equation}
D^n_k=\sum_{k'}D^{n-1}_{k'} (k'-1)[1-R_0^{-\lambda_{k'k}}] P(k|k') \prod_{m=0}^{n-1} \left(1-\frac{D^m_k}{V_k}\right)\,,
\label{eq:branching2}
\end{equation}
where each of the $D^{n-1}_{k}$	has $(k'-1)$ possible connections along which the infection can proceed ($- 1$ takes into account the link through which each of those subpopulations received the infection). In order to infect a subpopulation of degree $k$, three conditions need to occur: (i) the connections departing from nodes with degree $k'$ point to subpopulations of degree $k$, as indicated by the conditional probability $P(k|k')$; (ii) the reached subpopulations are not yet infected, as indicated by the probability $1-D^{n-1}_k/V_k$; (iii)
the outbreak seeded by $\lambda_{k'k}$	infectious individuals traveling from $k'$ to $k$ takes place, and the probability for this event to occur is given by $(1-R^{-\lambda_{k'k}}_0)$~\citep{Bailey1975}.
As done throughout the paper, we consider the case of uncorrelated networks in which the conditional
probability $P(k'|k)$ does not depend on the originating node, $P(k'|k)= k'P(k')/\langle k \rangle$~\citep{Barrat2008}.
Assuming that at the early stage of the epidemic only a few subpopulations are infected, i.e. $D^{n-1}_k / V_k \ll 1$ and the system is very close to the local epidemic threshold, i.e. $R_0 - 1 \ll 1$, we can further simplify Eq.~(\ref{eq:branching2}), by considering the series expansion:
\begin{equation}
(1-R^{-\lambda_{k'k}}_0) \simeq \lambda_{k'k}(R_0 -1)\,,
\label{eq:Rnot-series}
\end{equation}
which leads to the expression reported in Eq.~(\ref{eq:branching}).

In order to solve the Eq. ~(\ref{eq:branching}), we plug the explicit expression for the number of seeds: $\lambda_{k'k}=\alpha (\overline{N}_{kk'} + \overline{N}_{k'k})$, where the equilibrium populations are given by Eq.~(\ref{eq:pop0-deg}).
For the attack rate $\alpha$ we use the approximate relation valid for the condition $R_0 \approx 1$, $\alpha \approx 2(R_0-1)/R_0$~\citep{Murray2005}.
By replacing these variables in Eq. ~(\ref{eq:branching}), we find the relation:
\begin{multline}
D^n_k = C [ k^{\theta+1} \nu_k P(k) \sum_{k'}D^{n-1}_{k'} (k'-1)k'^{\theta+\chi} +\\+ k^{\theta+\chi+1}P(k)\sum_{k'}D^{n-1}_{k'}(k'-1)k'^{\theta} \nu_{k'} ]\,, 
\label{eq:iterative}
\end{multline}
where the constant $C$ is defined by:
\begin{equation}
C = \frac{2(R_0-1)^2}{R_0^2} \frac{\sigma \overline{N} \overline{\tau}}{\langle k \rangle \langle k^\phi \rangle \langle k^\chi \rangle}\,.
\label{eq:const}
\end{equation}
We can write a close form of the iterative process by defining the vector $\Theta^n = (\Theta^n_1, \Theta^n_2)$, whose components are:
\begin{eqnarray}
\Theta^n_1=\sum_k(k-1)k^{\theta+\chi}D_k^n\\
\Theta^n_2=\sum_k(k-1)k^\theta \nu_k D_k^n
\end{eqnarray}
The next generation equation can be written as $\Theta^n = C\,G\,\Theta^{n-1}$, with $G$ being the two dimensional matrix with elements:
\begin{eqnarray}
g_{11}=&g_{22}&=\langle (k-1)k^{2\theta+\chi+1} \nu_k \rangle \\
g_{12}&=&\langle (k-1)k^{2\theta+2\chi+1} \rangle \\
g_{21}&=&\langle (k-1)k^{2\theta+1} \nu_k^2 \rangle
\end{eqnarray}
The dynamical behavior of the system is determined by the largest eigenvalue of the matrix $G$, which is the quantity $\Lambda({P(k)}, \sigma, \overline{\tau}, \overline{N})$ defined by Eq.~(\ref{eq:lambda}), eventually leading to the expression for the global threshold $R_*$ of Eq.~(\ref{eq:threshold}).

If we assume that ill individuals do not leave home during their infectious period, but travel only to come back home if they were infected during their stay at destination, then the number of possible seeds from a subpopulation $k'$ to subpopulation $k$ is reduced to:
\begin{equation}
\lambda_{k'k}=\alpha \overline{N}_{k'k}\,.
\end{equation}
Under all the above mentioned assumptions, we can rewrite the iterative process described by Eq.~(\ref{eq:iterative}) as:
\begin{equation}
D^n_k = C k^{\theta+\chi+1}P(k)\sum_{k'}D^{n-1}_{k'}(k'-1)k'^{\theta} \nu_{k'}\,, 
\end{equation}
where the constant $C$ is still defined by Eq.~(\ref{eq:const}).
By defining the variable $\Theta^n=\sum_k(k-1)k^\theta \nu_k D_k^n$, the next generation equation can be written as:
\begin{equation}
\Theta^n = C  \langle (k-1)k^{2\theta+\chi+1} \nu_k \rangle \Theta^{n-1} \,.
\end{equation}
from which the global threshold condition of Eq.~(\ref{eq:th_1seed}) is immediately derived.

\section{Global invasion threshold for the homogeneous network \label{app:homo_threshold}}
Here, we derive the global invasion threshold parameter $R_*$ for the case of a homogeneous network. 
All nodes have the same degree $\overline{k}$ and the same population $\overline{N}$. The mobility fluxes along the links are homogeneously distributed and are determined by the leaving rate $\sigma_{kk'}= \sigma \overline{k}^{2 \theta}$ and the length of stay $\tau_k = \overline{\tau}$. The invasion process is described by the dynamics of the diseased subpopulations at generation $n$, $D^n$, that is governed by the equation
\begin{equation}
D^n = D^{n-1}(\bar k-1) \left(1-R_0^{-\lambda_{\bar k \bar k}}\right) \left(1-  \frac{D^{n-1}}{V}\right),
\label{eq:Dn}
\end{equation}
that is the analogous of Eq.~(\ref{eq:branching2}). The number of seeds, $\lambda_{\bar k \bar k}$, is given by the expression
\begin{equation}
\lambda_{\bar k \bar k}= 2 \alpha  \, \sigma\bar N \bar \tau  \; \bar k^{2 \theta}  \nu_{\bar k},
\label{eq:seeder}
\end{equation}
where $\alpha$ is the attack rate of the SIR epidemic and $\nu_{\bar k}=\left ( 1+ \sigma \bar \tau \;   \bar k^{2 \theta +1} \right) ^{-1} $. 
We plug Eq.~(\ref{eq:seeder}) into Eq.~(\ref{eq:Dn}), and we assume that the epidemic is at the early stage (thus $1- {D^{n-1}}/{V} \simeq 1$), and that $R_0$ is close to unit (thus $\alpha \simeq {2(R_0-1)}/{R_0^2}$~\citep{Murray2005}). Then, we recover an explicit form for Eq.~(\ref{eq:Dn}):
\begin{equation}
D^n = 4  \sigma \bar N \bar \tau \; \frac{(R_0-1)^2}{R_0^2} \; \bar k^{2 \theta} (\bar k-1) \nu_{\bar k} D^{n-1}\,,
\label{eq:Dn_espl}
\end{equation}
which directly provides the definition of $R_*$. Indeed, $D^n$ will grow exponentially and the epidemic will invade a non-infinitesimal fraction of the network if the following threshold condition is satisfied:
\begin{equation}
R_* \equiv 4  \sigma \bar N \bar \tau \; \frac{(R_0-1)^2}{R_0^2} \; \bar k^{2 \theta} (\bar k-1) \nu_{\bar k} >1.
\label{eq:thresh_hom2}
\end{equation}
When sick individuals change their travel behavior and do not leave home, the number of possible seeds is simply reduced by a factor $2$ since the infection can arrive from one source only:
\begin{equation}
\lambda_{\bar k \bar k}= \alpha  \, \sigma\bar N \bar \tau  \; \bar k^{2 \theta}  \nu_{\bar k}.
\label{eq:1seeder}
\end{equation}
Under the same assumptions described above, the final equation for the global invasion threshold changes to:
\begin{equation}
R_* = 2  \sigma \bar N \bar \tau \; \frac{(R_0-1)^2}{R_0^2} \; \bar k^{2 \theta} (\bar k-1) \nu_{\bar k}.
\label{eq:thresh_hom1seed}
\end{equation}

\section{Global disease dynamics above the threshold \label{app:global_prevalence}}
The infection evolution on a metapopulation system characterized by a non-Markovian dynamics can be formalized as a deterministic process ruled by non linear differential equations which describe the time evolution of each compartment, for each of the subclasses $X_{ii}$ and $X_{ij}$ (where $X=S, I$ or $R$) and in each subpopulation $i$.
The reaction-diffusion epidemic dynamics is encoded in the rate equations describing the coupling between infection transmission and traveling. 
For the infectious compartments $I_{ii}$ and $I_{ij}$ such equations read:
\begin{equation}
\begin{aligned}
d_t I_{ii}(t) &= S_{ii} (t) \mathcal F_i(t) - \mu I_{ii}(t) - \sigma_i I_{ii}(t) + \sum_{l \in \upsilon(i)} \frac 1 {\tau_{l}} I_{il}(t)\\
d_t I_{ij}(t) &= S_{ij} (t) \mathcal F_j(t) - \mu I_{ij}(t) + \sigma_{ij} I_{ii}(t) -  \frac 1 {\tau_{j}} I_{ij}(t),  
\end{aligned}
\label{eq:global1}
\end{equation}
where $\mathcal F_i(t)$ is the force of infection in the subpopulation $i$, given by:
\begin{equation}
\mathcal F_i(t)= \beta I_i^*(t)/N_i^*(t)\,,
\end{equation}
where, by definition:
\begin{eqnarray}
S^*_i(t) &\equiv&  S_{ii} (t) + \sum_{j \in \upsilon(i)} S_{ji} (t)\\ 
I^*_i(t) &\equiv&  I_{ii} (t) + \sum_{j \in \upsilon(i)} I_{ji} (t)\\ 
R^*_i(t) &\equiv&  R_{ii} (t) + \sum_{j \in \upsilon(i)} R_{ji} (t)\\ 
N^*_i(t) &\equiv& S^*_i(t) + I^*_i(t) + R^*_i(t)\,.
\end{eqnarray} 
The system of rate equations Eq.~(\ref{eq:global1}) can be combined into a single equation giving the evolution of the infected individuals resident in the subpopulation $i$:
\begin{equation}
 d_t I_{i}(t) =  S_{ii} (t) \mathcal F_i(t) - \sum_{j} S_{ij} (t) \mathcal F_j(t) - \mu I_i(t) \,, 
\end{equation}
which allows the computation of the dynamical equation for the total number of infectious $I(t)$ in the metapopulation system
\begin{equation}
d_t I(t) = \sum_i d_t I_{i}(t) = \sum_i S_{ii} (t) \mathcal F_i(t) - \sum_{ij} S_{ij} (t) \mathcal F_j(t) - \mu I(t) \,.
\end{equation}
We can rename the indexes of the second sum in the r.h.s. of the above equation, $\sum_{ij} S_{ij} (t) \mathcal F_j(t) \equiv \sum_{ji} S_{ji} (t) \mathcal F_i(t)$, and finally obtain the Eq.~(\ref{eq:global_prevalence}):
\begin{equation}
\frac{dI(t)}{dt} = \beta \sum_{i \in V} S^*_i(t)\frac{I^*_i(t)}{N^*_i(t)} - \mu I(t) \,.
\end{equation}
that describes the time evolution of the total number of infectious individuals of the system.

\bibliographystyle{model2-names.bst}
\bibliography{references.bib}

\end{document}